\documentclass[prd,aps,showpacs,twocolumn,superscriptaddress,floatfix,nofootinbib,preprintnumbers]{revtex4-1}

\usepackage{graphicx} %
\usepackage[dvipsnames]{xcolor}
\usepackage{appendix}
\usepackage{amsmath,latexsym}
\usepackage{mathtools}
\usepackage{combelow}
\usepackage{ulem}
\usepackage[bookmarks=true, colorlinks=true, linkcolor=RoyalBlue, urlcolor=blue, citecolor=OliveGreen, pdftex, bookmarks=true, hyperindex=true]{hyperref}

\renewcommand{\d}{\mathrm{d}}
\newcommand{\xT}{\mathbf{x_\perp}}

\newcommand{\trento}{T$_{\rm R}$ENTo~}

\begin{document}
\title{Elliptic and triangular flow responses in O$+$O and Ne$+$Ne collisions from hydrodynamics and kinetic theory}
\author{Victor E. Ambru\cb{s}}
\affiliation{%
Fakultät für Physik, Universität Bielefeld, D-33615 Bielefeld, Germany
}
\affiliation{Department of Physics, West University of Timi\cb{s}oara, \\
Bd.~Vasile P\^arvan 4, Timi\cb{s}oara 300223, Romania}
\author{Giuliano Giacalone}
\affiliation{Theoretical Physics Department, CERN, CH-1211 Gen\`eve 23, Switzerland}
\author{Clemens Werthmann}
\email{clemens.werthmann@ugent.be}
\affiliation{Department of Physics and Astronomy, Ghent University, 9000 Ghent, Belgium}
\author{Fabian Zhou}
\affiliation{Institute for Theoretical Physics, University of Heidelberg, 69120 Heidelberg, Germany}
\affiliation{Instituto Galego de Física de Altas Enerxías (IGFAE), Universidade de Santiago de Compostela
E-15782 Galicia, Spain}
\date{\today}

\preprint{CERN-TH-2026-276}

\begin{abstract}
Recent light-ion collisions at the Large Hadron Collider provide an opportunity to investigate the system-size dependence of hydrodynamization and whether the structure of light nuclei can be extracted from collective flow data. Both objectives require an accurate understanding of how the collision dynamics converts initial-state geometry into final-state momentum anisotropies. To this end, and to assess the impact of the experimental results, we compare simulations of O+O and Ne+Ne collisions in hydrodynamic and kinetic theory simulations across opacities. In both ultracentral and peripheral collisions, Ne+Ne-to-O+O elliptic flow ratios are robust against changes in dynamical description, supporting interpretations of data based on nuclear-geometry-driven effects. Triangular-flow ratios instead show substantially stronger sensitivity to nonequilibrium dynamics.
\end{abstract}

\maketitle

\section{Introduction}

The hydrodynamic paradigm of heavy-ion collisions has been established primarily through its successful description of collective flow, particularly anisotropic flow~\cite{Kolb:2003dz,Teaney:2009qa,Heinz:2013th,Hirano:2012kj,Florkowski:2017olj,Song:2017wtw,Heinz:2024jwu}. These observables have enabled quantitative constraints on the dynamical properties of the created medium and the initial conditions of its evolution, notably through comprehensive frameworks of global Bayesian inference~\cite{Putschke:2019yrg,Bernhard:2019bmu,JETSCAPE:2020shq,Nijs:2020roc}. However, the observation of similar collective phenomena in high-multiplicity proton-proton and proton-nucleus collisions has challenged the expected limits of applicability of this paradigm, suggesting the formation of a medium with collective behavior resembling that observed in nucleus-nucleus collisions~\cite{Grosse-Oetringhaus:2024bwr}. Characterizing the matter produced in these small systems has thus emerged as one of the central goals of high-energy nuclear physics and the study of out-of-equilibrium QCD matter \cite{Berges:2020fwq}.

In parallel, the possibility of probing the initial states has also received a lot of attention in recent years \cite{Jia:2022ozr}, as they encode information on the spatial structure of the collided nuclei, opening a new experimental window onto low-energy nuclear structure and correlations \cite{Giacalone:2023hwk,Blaizot:2025scr,Duguet:2025hwi,Blaizot:2025bfu,Bofos:2026huw,Mehrabpour:2026yuc,Bofos:2026nmg,Ding:2026foi,Deng:2026wxb,Barata:2026aak}, and associated phenomena such as deformation \cite{Zhang:2021kxj,Xu:2021uar,Nijs:2021kvn,Ryssens:2023fkv,Zhao:2024lpc,Fortier:2023xxy,Giacalone:2024ixe,Mantysaari:2024uwn,Xu:2025cgx,Ke:2025tyv,Mehrabpour:2026lhj} and clustering \cite{Zhao:2024feh,Liu:2025zsi,Li:2025hae,Mehrabpour:2025ogw,MenonKavumpadikkalRadhakrishnan:2026tfn}. These features can be optimally probed through observable ratios between two collision systems of similar size~\cite{STAR:2021mii,Giacalone:2025vxa}, or through the correlation of anisotropic flow coefficients with the mean transverse momentum~\cite{Giacalone:2019pca,Giacalone:2020awm,Jia:2021wbq,Bally:2021qys,Jia:2021qyu,ATLAS:2022dov,Samanta:2023qem,Bally:2023dxi,Gardim:2023ksn,Fortier:2024yxs,Wang:2024vjf,STAR:2025elk,Hagino:2025vxe,Giacalone:2026fat}. Although modern Bayesian analysis frameworks prove capable of inferring initial-state properties from final-state data when dedicated observables are fitted~\cite{Giacalone:2023cet,Giacalone:2026fat}, a precise extraction requires an adequate level of understanding of the evolution dynamics at play, especially when small systems are considered.

In view of this, results on anisotropic flow results in light-ion collisions involving $^{16}$O and $^{20}$Ne nuclei  \cite{ATLAS:2025nnt,ALICE:2025luc,CMS:2025tga,ATLAS:2026zgq}, recently performed at the Large Hadron Collider (LHC), open opportunities on two fronts \cite{Giacalone:2024luz}. These collisions allow us to probe both the degree of hydrodynamization of small system sizes as well as the capability of contemporary frameworks in extracting nuclear structure properties in systems that are more dominated by fluctuations than those in heavy-ion collisions. Light-ion collisions are essentially understood as lying at the border between dilute transport and hydrodynamic regimes \cite{Ambrus:2022koq,Ambrus:2022qya}. However, inaccuracies in dynamical models do not necessarily obscure nuclear structure information. They may, for example, cancel in appropriate combinations of flow observables. Thus, a detailed study of this interplay is needed. 

In this paper, we investigate this by combining advanced nuclear-structure modeling with kinetic-theory simulations. To test the applicability of hydrodynamics, we compare hydrodynamic and kinetic-theory predictions for light-ion collisions across a broad range of opacities, from dilute transport toward the ideal-hydrodynamic limit. We employ the conformal relaxation time approximation (RTA) framework of Refs.~\cite{Ambrus:2022koq,Ambrus:2022qya,Ambrus:2024hks,Ambrus:2024eqa}. RTA is also employed within the CoMBolt-RTA framework~\cite{Taghavi:2025xhl,Taghavi:2025ddm}, which was recently extended to nonconformal systems~\cite{Taghavi:2026vyw}. Regarding nuclear structure, we use oxygen and neon inputs calculated in \cite{Giacalone:2024luz} from the \textit{ab initio} Projected Generator Coordinate Method (PGCM)~\cite{Frosini:2021tuj,Frosini:2021sxj,Frosini:2021ddm,Frosini:2021fjf}. They allow us to investigate how nuclear properties, such as the prolate deformation of $^{20}$Ne or the octupole deformation of $^{16}$O, encoding possible $\alpha$-clustering effects, influence the subsequent dynamics.

We examine thus several aspects of the elliptic and triangular flow responses in kinetic theory and hydrodynamic simulations of Ne+Ne and O+O collisions. We quantify to which extent hydrodynamics is an appropriate proxy for full kinetic theory calculations on both absolute magnitudes of flow coefficients and on ratios taken between light-ion systems. We study how flow responses and nuclear-structure influences play out as a function of interaction strength, and also test the effects of inhomogeneous cooling before transverse expansion. In order to include triangular flow in this analysis, we introduce a proxy that allows directly extracting it from a hydrodynamic description without the need for particlization.

In the following, Sec.~\ref{sec:2} introduces the kinetic theory and hydrodynamic frameworks employed in our analysis, as well as the observables of interest. Section~\ref{sec:3} discusses features of the initial conditions of our simulations and their dependence on opacity. Section~\ref{sec:4} then presents our results from the dynamical simulation of elliptic and triangular flow responses, as well as their interplay with nuclear deformation effects. Section~\ref{sec:5} contains concluding remarks and an outlook. Further technical details on formulations, derivations, and additional figures are relegated to the various appendices.

\section{Theoretical Setup}
\label{sec:2}

Subsections~\ref{sec:kinetic_theory}, \ref{sec:loebe} and \ref{sec:hydro} describe the kinetic theory setup, its leading-order (single-hit) approximation, and the hydrodynamic setup. Sec.~\ref{sec:init} describes the initial state, while in Sec.~\ref{sec:obs} we introduce our observables.

\subsection{Kinetic theory}
\label{sec:kinetic_theory}

For all results from kinetic theory based simulations in this paper, we employ the relativistic Boltzmann equation in the Anderson-Witting relaxation-time approximation \cite{Anderson:1974,Anderson:1974b},
\begin{equation}
 p^\mu \partial_\mu f = C_{\rm AW} = -\frac{ p^\mu u_\mu}{\tau_R} (f - f_{\rm eq}),
 \label{eq:boltz}
\end{equation}
where the relaxation time $\tau_R$ is fixed by imposing a constant shear viscosity to entropy density ratio $\eta / s$,
\begin{equation}\label{eq:relaxation_time}
 \tau_R = \frac{5 \eta}{sT},
\end{equation}
where $\eta = \frac{4}{5} \tau_R P$ and $s = 4P / T$ for massless particles. The equilibrium distribution is taken as the Bose-Einstein distribution,
\begin{equation}
 f_{\rm eq} = \frac{1}{e^{u_\mu p^\mu / T} -1},
\end{equation}
where the four-velocity $u^\mu$ and the local temperature $T$ are obtained via Landau matching, by imposing energy-momentum conservation:
\begin{equation}
 T^{\mu}{}_\nu u^\nu = \epsilon u^\mu, \quad 
 \epsilon = aT^4, \quad a = \frac{\pi^2}{30} \nu_{\rm eff},
\end{equation}
with the effective number of bosonic degrees of freedom $\nu_{\rm eff} = 42.25$ corresponding to $8$ gluons and $2.5$ massless quark flavors, compatible with high-temperature lattice QCD results \cite{HotQCD:2014kol,Borsanyi:2016ksw}. The energy-momentum tensor $T^{\mu\nu}$ is given in terms of the phase space distribution $f$ as
\begin{equation}
 T^{\mu\nu} = \int dP p^\mu p^\nu f, \label{eq:Tmunu}
\end{equation}
where the generally covariant momentum integration element is given by \cite{Cercignani:2002}
\begin{equation}
 dP = \nu_{\rm eff} \sqrt{-g} \frac{d^3p}{(2\pi)^3 p_0},
\end{equation}
with $\sqrt{-g}$ representing the square root of the metric determinant.

In this paper, we consider a longitudinally boost invariant system, expanding along the $z$ axis, and study its dynamics in the transverse ($xy$) plane. It is thus convenient to switch from Minkowski to Milne coordinates $(t,\mathbf{x}_\perp,z)\leftrightarrow (\tau,\mathbf{x}_\perp ,\eta)$ where the Bjorken time $\tau$ and the spacetime rapidity $\eta$ are defined by
\begin{equation}\label{eq:Milne_coordinates}
 \tau = \sqrt{t^2 - z^2}, \quad 
 \tanh \eta = \frac{z}{t}.
\end{equation}
Analogously, we introduce the momentum rapidity $y$ via
\begin{equation}
 \tanh y = \frac{p^z}{p^t}.
 \end{equation}
Boost invariance dictates that the distribution function depends on $\eta$ and $y$ only through their difference, $y -\eta$. Parametrizing the momentum space with respect to\newline $(\mathbf{p}_\perp, y - \eta)$, the Boltzmann eq.~\eqref{eq:boltz} becomes
\begin{equation}
 p^\tau \frac{\partial f}{\partial \tau} + \mathbf{p}_\perp \cdot \boldsymbol{\nabla} f - p^\eta \frac{\partial f}{\partial(y - \eta)} = -\frac{p \cdot u}{\tau_R}(f - f_{\rm eq}),
 \label{eq:boltz_y}
\end{equation}
with $p^\tau = p_\perp \cosh(y - \eta)$ and $\tau p^\eta = p_\perp \sinh(y - \eta)$.
A more detailed description of employing the Milne coordinates is given in App.~\ref{Milne_coordinates}.

In principle, we seek to study the system evolution during the full preequilibrium phase, hence the initialization time $\tau_0$ must be pushed to $0$. In this case, the particle distribution function becomes squeezed to the early time limit of the hydrodynamic attractor in kinetic theory \cite{Ambrus:2021fej}
\begin{equation}
 f_0(\mathbf{x}_\perp, \mathbf{p}_\perp, y - \eta) \to \frac{(2\pi)^3}{\nu_{\rm eff}} \frac{\delta(y -\eta)}{\tau_0 p_\perp} \frac{dN_0}{d^2\mathbf{x}_\perp d^2\mathbf{p}_\perp dy}\,,\label{eq:f0}
\end{equation}
where we assume that $dN_0 / d^2\mathbf{x}_\perp d^2 \mathbf{p}_\perp dy$ is independent of the momentum azimuth, $\varphi_p = \arctan(p^y / p^x)$. In this case, the energy-momentum tensor at initial time takes the form
\begin{equation}
 T^{\mu\nu}_0 = {\rm diag}(\epsilon_0, P_{T;0}, P_{T;0}, P_{L;0}),
\end{equation}
with $P_{L;0}= 0$ and $P_{T;0} = \epsilon_0 / 2$, while the initial energy density is related to the particle distribution via
\begin{equation}\label{eq:init_e_dens}
 \tau_0 \epsilon_0 = \int d^2\mathbf{p}_\perp p_\perp\frac{dN_0}{d^2\mathbf{x}_\perp d^2\mathbf{p}_\perp dy}.
\end{equation}
No further details need to be specified regarding the initial particle distribution, as our focus in this paper is entirely on energy-weighted observables (see Sec.~\ref{sec:obs}). In our numerical simulations, we considered a finite initialization time $\tau_0 = 10^{-6} R$ and a finite longitudinal pressure $P_{L;0} \neq 0$ was introduced to avoid the singular structure implied by the delta function in Eq.~\eqref{eq:f0} (see Appendix~\ref{numerical_solution} for details).

The total energy available in the system at initial time can be described by the energy per unit rapidity, 
\begin{equation}
 \frac{dE_0}{d\eta} = \int d^2\mathbf{x}_\perp\, \tau_0 \epsilon_0\,,
\end{equation}
which at early times is entirely in transverse energy, see Eq.~\eqref{eq:f0}. We further characterize the initial energy distribution by the transverse system size $R$, given by
\begin{equation}
 R^2 \frac{dE_0}{d\eta} = \int d^2\mathbf{x}_\perp\, \tau_0 \epsilon_0 \mathbf{x}_\perp^2.
\end{equation}

Using these global quantities, the Eq. \eqref{eq:boltz_y} can now be expressed in a dimensionless way. This allows us to trade the relaxation time $\tau_R$, governing the evolution described by the Boltzmann Eq.~\eqref{eq:boltz}, for a dimensionless parameter called the opacity $\hat{\gamma}$~\cite{Kurkela:2019kip},
\begin{equation}
    \hat{\gamma} = \frac{1}{5\eta/s} \left(\frac{R}{\pi a} \frac{dE_0}{d\eta}\right)^{1/4},
\end{equation}
which characterizes roughly how strongly interacting the system is. For small systems like proton-proton one expects $\hat{\gamma}\sim1$ whereas large system like Pb+Pb are much denser, i.e. $\hat{\gamma}\gg 1$ (for LHC energies). Thus, it solely controls the degree of relaxation towards equilibrium which is determined by the overall energy scale $dE_0 / d\eta$, system size $R$, partonic degrees of freedom (through the parameter $a$) and shear viscosity to entropy density ratio $\eta / s$.  In this work, we will consider $4\pi \eta / s = 1.5$ (compatible with Bayesian analysis estimates \cite{Bernhard:2019bmu}) and $5$ (a more weakly-interacting regime). 

Lastly, we can facilitate the computation by integrating out the total momentum degree of freedom. Doing this for Eq.~\eqref{eq:boltz_y} decouples the energy weighted reduced distribution function $\mathcal{F}$, defined via
\begin{equation}
 \mathcal{F} = \frac{\nu_{\rm eff}}{(2\pi)^3} \int_0^\infty dp^\tau\, (p^\tau)^3 f, \label{eq:F_red}
\end{equation}
from other degrees of freedom, owed to the simple form of the RTA collision kernel. This is the function that we solve for in the numerical setup.
Accordingly, the initial and equilibrium reduced distributions, respectively, are given by
\begin{equation}
    \mathcal{F}_0 = 
 \frac{\epsilon_0}{2\pi} \delta(v^z), \quad 
 \mathcal{F}_{\rm eq} = \frac{\epsilon}{4\pi (v^\mu u_\mu)^4}.
 \label{eq:Feq_red}
\end{equation}
More details on the numerical setup are described in App.~\ref{numerical_solution}, including the exact equation that is solved numerically, which is Eq.~\eqref{eq:boltz_reduced}.

\subsection{Linear order in opacity}\label{sec:loebe}
For kinetic theory at small interaction rates, the solution of the Boltzmann equation can be expanded in the number of scatterings per particle. This expansion scheme was introduced in Refs.~\cite{Heiselberg:1998es,Borghini:2010hy} and has more recently also been used in Refs.~\cite{Romatschke:2018wgi,Kurkela:2018ygx,Borghini:2018xum,Kurkela:2021ctp, Ambrus:2021fej,Bachmann:2022cls}. The zeroth order result $f^{(0)}$ is the free-streaming case 
\begin{align}
    p^\mu\partial_\mu f^{(0)}=0\,.
\end{align}
The linear order correction $f^{(1)}$ is then obtained by computing the effect of scatterings based on the zeroth order solution,
\begin{align}
    p^\mu\partial_\mu f^{(1)}=C[f^{(0)}]\,.
\end{align}
This turns the right-hand side into a source term instead of a functional of the unknown function, such that the equation can be solved by integration. In the case of RTA kinetic theory, the expansion parameter is the opacity $\hat{\gamma}$ \eqref{eq:ghat}. We use the previously developed code LOebe to obtain event-by-event final state results based on this scheme. For a description of this code, please see Refs.~\cite[App. B]{Ambrus:2022koq} and~\cite[App. A]{Ambrus:2024hks}.

\subsection{Hydrodynamics}\label{sec:hydro}

The dynamics of a system not far from equilibrium can be described by the theory of relativistic hydrodynamics. In the Landau frame, the energy-momentum tensor $T^{\mu\nu}$ of a conformal fluid can be decomposed with respect to the fluid four-velocity $u^\mu$ as follows:
\begin{equation}
 T^{\mu\nu} = \epsilon u^\mu u^\nu - P \Delta^{\mu\nu} + \pi^{\mu\nu},
\end{equation}
where $\pi^{\mu\nu}$ represents the shear-stress tensor and $\Delta^{\mu\nu} = g^{\mu\nu} - u^\mu u^\nu$. Imposing the conservation law $\partial_\mu T^{\mu\nu} = 0$ leads to the energy and momentum equations,
\begin{align}
 D \epsilon + (\epsilon + P) \theta - \pi^{\mu\nu} \sigma_{\mu\nu} &= 0, \nonumber\\ 
 (\epsilon +P) Du^\mu - \nabla^\mu P + \Delta^\mu_\lambda \partial_\nu \pi^{\lambda\nu} &=0, 
 \label{eq:hydro_cons}
\end{align}
where $D = u^\mu \partial_\mu$ is the comoving (material) derivative, $\theta = \partial_\mu u^\mu$ is the fluid expansion scalar, $\nabla^\mu = \Delta^{\mu\lambda} \partial_\lambda$, $\sigma_{\mu\nu} =  \partial_{\langle \mu} u_{\nu \rangle}$ is the shear tensor and $A_{\langle \mu\nu \rangle} = \Delta^{\alpha\beta}_{\mu\nu} A_{\alpha\beta}$ with $\Delta^{\alpha\beta}_{\mu\nu} = \frac{1}{2}(\Delta^\alpha_\mu \Delta^\beta_\nu + \Delta^\beta_\mu \Delta^\alpha_\nu) - \frac{1}{3} \Delta^{\alpha\beta} \Delta_{\mu\nu}$. 

To close the system of Eqs.~\eqref{eq:hydro_cons}, we employ the M\"uller-Israel-Stewart second-order hydrodynamics theory \cite{Muller.1967,Israel.1979}, which models the evolution of the shear-stress tensor $\pi^{\mu\nu}$ via
\begin{multline}
 \tau_\pi D \pi^{\langle \mu\nu \rangle} + \pi^{\mu\nu} = 2\eta \sigma^{\mu\nu} + 2\tau_\pi \pi^{\langle \mu}_\lambda \omega^{\nu\rangle \lambda} -\delta_{\pi\pi} \pi^{\mu\nu} \theta \\
 - \tau_{\pi\pi} \pi^{\lambda \langle \mu} \sigma^{\nu\rangle}_\lambda + \phi_7 \pi^{\langle \mu}_\alpha \pi^{\nu \rangle \alpha},
 \label{eq:hydro_pi}
\end{multline}
where $\omega_{\mu\nu} = \frac{1}{2} (\nabla_\mu u_\nu - \nabla_\nu u_\mu)$ is the vorticity tensor. The quantities $\tau_\pi$, $\eta$, $\delta_{\pi\pi}$, $\tau_{\pi\pi}$ and $\phi_7$ represent transport coefficients that need to be provided as external input, e.g. from a more microscopic theory. To achieve a meaningful comparison between the RTA kinetic theory described in Sec.~\ref{sec:kinetic_theory} and hydrodynamics, we take these transport coefficients to match those corresponding to the RTA system, namely \cite{Jaiswal:2013npa,Molnar:2013lta,Ambrus:2022vif}:
\begin{align}
 \eta = \frac{4}{5} \tau_\pi P, \quad
 \delta_{\pi\pi} = \frac{4\tau_\pi}{3}, \quad 
 \tau_{\pi\pi} = \frac{10\tau_\pi}{7}, \quad
 \phi_7 = 0,
 \label{eq:tcoeffs}
\end{align}
with $\tau_\pi = \tau_R$.

We solve Eqs.~\eqref{eq:hydro_cons} and \eqref{eq:hydro_pi} numerically using the viscous HLLE (vHLLE) code \cite{Karpenko:2013wva},\footnote{Commit number \texttt{efa9e28d24d5115a8d813485232fb342b38380f0}.} which we adapted to the setup described in this paper (see Refs.~\cite{Ambrus:2021fej,Ambrus:2022koq} for a more detailed discussion). The simulation setup is similar to the one described in App.~\ref{numerical_solution}, but we took a larger spatial grid  employing $401 \times 401$ nodes, since advection is implemented in vHLLE using only a second-order scheme.

\subsection{Initial energy density}\label{sec:init}

We employ \trento\cite{Moreland:2014oya} to generate event-by-event initial profiles of the energy density. The output of \trento is interpreted as an energy density, 
\begin{align}
    \tau_0\epsilon(\tau_0,\xT)=N E_{\rm ref}^{2-2q} \left\{ \frac{1}{2}\left[T_A(\xT)^p+T_B(\xT)^p\right]\right\}^{q/p}\,,
\end{align}
where $E_{\rm ref}=0.2$ GeV and $T_A$ and $T_B$ are the local thickness functions of the colliding nuclei evaluated at transverse location ${\bf x_\perp}$.
In order to accurately account for nuclear structure effects, particularly nuclear deformations, we use sampled nucleon positions in the $^{20}$Ne and $^{16}$O nuclei from the PGCM model of Ref.~\cite{Giacalone:2024luz}. For the \trento parameters, we use the MAP values from Bayesian analyses in Trajectum~\cite{Nijs:2025qxm,Nijs:2023yab}, except for the number of nucleon constituents, which we set to $n_c=3$ for simplicity. For definiteness, the values we use are
\begin{align}
   & \sigma=6.83\,{\rm fm}^2\,,~ 
N=22.9\,{\rm GeV}, ~
p=-0.079\,, \nonumber\\
&q=1.25\,, ~
 k=1.28\,, ~
w=0.578\,{\rm fm}\,, ~
v=0.5\,{\rm fm} \,. 
\end{align}
In order to map out the properties of this ensemble of initial states, we generate 51.2 million events for both O+O and Ne+Ne collisions, and group them into centrality classes based on an estimation formula for the final-state charged-particle multiplicity at midrapidity~\cite{Giacalone:2019ldn}. For the simulations, we use subsets of 51\,200 events each, from either the 0-1\% ultracentral class or the 45-46\% peripheral class (chosen as reference for peripheral events). We also pick 19\,340 events from the 0-1\% centrality class of Ne+Ne collisions in which the orientation angle $\theta$ to the beam axis of both nuclei fell into the same of 20 equally sized bins in $|\cos(\theta)|$. In the PGCM calculation, the symmetry breaking direction is chosen as the $z$-axis, so this is the direction in which the tip of the prolate neon nucleus originally points. We then rotate the nuclei and keep track of $\theta$.

In order to fix the initial state in kinetic theory, we also need to prescribe the momentum space distribution. We choose to initialize on the early time attractor, i.e. with (close to) zero longitudinal momenta, and no anisotropy in transverse momentum. Accordingly, we initialize at early times $\tau_0=10^{-6}R$, where the system is free-streaming and $\tau \epsilon$ is constant in time, such that results do not depend on the precise choice of $\tau_0$. The distribution in the magnitude of momentum need not be defined as we only consider moments of the distribution function with dimension of energy. 

As we want to meaningfully compare hydrodynamic results to kinetic theory, we need a setup in which both descriptions are in agreement at hydrodynamization time, as the theories predict deviating evolutions during the initial period. Most notably, hydrodynamics features a negative longitudinal pressure at very early times, which leads to an increase in $\tau\epsilon$.
%One way to do this is to employ a hybrid scheme, switching from kinetic theory to hydrodynamics. This, however, requires to make a modeling choice for when to switch. 
As explained in detail in Sec. IV D of Ref.~\cite{Ambrus:2022koq}, we locally scale the initial energy density of hydrodynamics to counteract the early time difference to kinetic theory, as predicted via the local Bjorken flow attractor. The initialization time of the hydrodynamic simulations is taken as $\tau_0=10^{-3}R$.

\subsection{Observables}\label{sec:obs}

We are interested in elliptic and triangular flow. They depend on the initial state properties that determine the number of interactions in the system through the opacity, which was already introduced in Sect.~\ref{sec:kinetic_theory}. More crucially, these anisotropic flow observables are generated through the medium response to the initial transverse anisotropies in position space. These are quantified through eccentricities, $\epsilon_n$, defined by \cite{Teaney:2010vd,Sousa:2024msh}
\begin{align}
    \epsilon_ne^{in\Psi_n}=-\frac{\int_\xT |\xT|^ne^{in\varphi_x}\,\epsilon(\tau_0,\xT)}{\int_\xT |\xT|^n\,\epsilon(\tau_0,\xT)}\,,
\end{align}
%
% \begin{align}
%     \epsilon_2e^{2i\Psi_2}=\frac{\int_\xT |\xT|^2e^{2i\varphi_x}\,e(\tau_0,\xT)}{\int_\xT |\xT|^2\,e(\tau_0,\xT)}
% \end{align}
%
% and the triangularity $\epsilon_3$ defined through
%
% \begin{align}
%     \epsilon_3e^{3i\Psi_3}=\frac{\int_\xT |\xT|^3e^{3i\varphi_x}\,e(\tau_0,\xT)}{\int_\xT |\xT|^3\,e(\tau_0,\xT)}\,,
% \end{align}
where the transverse coordinates $\xT$ have been centered to the center of mass of the density profile and $\varphi_x$ is the azimuthal angle in transverse position space. The final-state flow anisotropies are instead quantified in terms of Fourier coefficients $v_n$ in the particle distribution in transverse momentum, defined as
\begin{align}
    v_ne^{in\Psi_{p,n}}=\frac{\int \d P\int_\xT p^0 p_\perp e^{in\varphi_p}\,f}{\int \d P\int_\xT p^0p_\perp\,f}\,,\label{eq:vn_def}
\end{align}
where $\varphi_p$ is the azimuthal angle in transverse momentum space. 
As in our earlier works, we add an additional weight of the particles' transverse momenta (see also Ref.~\cite{Peng:2026qxz}). These modified flow observables have been shown to deviate from the usual ones by a $\sqrt{s_{NN}}$-dependent conversion factor~\cite{Kurkela:2019kip}. 

Now, to measure the momentum anisotropy directly in hydrodynamics, we consider proxies of these $v_n$ observables that can be computed directly from the energy-momentum tensor. This circumvents ambiguities in the particlization procedure. For elliptic flow, the corresponding observable $\varepsilon_p$ is defined as
\begin{align}
    \varepsilon_pe^{2i\Psi_{p,2} }&=\frac{\int_\xT T^{xx}-T^{yy} + 2i T^{xy}}{\int_\xT T^{xx}+T^{yy} }\\
    &=\frac{\int\d P \int_\xT p_\perp^2\, e^{2i\varphi_p}\,f }{\int \d P \int_\xT p_\perp^2 \,f}\,.
\end{align}
The second line shows that in kinetic theory, this definition is very similar to that of $v_2$. 

For the flow triangularity coefficient $v_3$, it is less straightforward to construct such a proxy, as in contrast to the elliptic case, the energy-momentum tensor itself does not have any components that correspond to weights of threefold rotational symmetry. Instead, we construct an approximation to $v_3$ by employing a linearization around equilibrium, $f\approx f_{\rm eq}+\delta f$, into Eq.~\eqref{eq:vn_def}. The details of this derivation are laid out in App.~\ref{sec:momentum_triangularity}. This leads to

\begin{widetext}
    \begin{align}
     \varepsilon_{p,3}e^{3i\Psi_{p,3}}=\frac{\int_\xT\frac{4 \gamma \epsilon}{3} e^{3i \varphi_u} \beta^3 \gamma +\frac{1}{2} \left[\beta e^{i \varphi_u} \pi_{(2)} (6 - 3\beta^2 + \beta^4) -
 \beta^3 e^{3i \varphi_u} \pi_{(0)} (5 - 2 \beta^2)
 + \beta^5 e^{5i \varphi_u}\pi_{(-2)} \right]}{\int_\xT\frac{\epsilon}{6} Q_1(\beta)+\frac{1}{16 \beta^2 \gamma^2} \left[\pi_{(0)} Q_2(\beta)  - \frac{\pi^{00}}{\beta^2} Q_3(\beta) \right]}\,,\label{eq:ep3}
\end{align}
\end{widetext}
where $\gamma=u^0$, $\beta=\sqrt{-u^iu_i}/u^0$ and $\varphi_u=\arctan(u^y/u^x)$ are related to the flow velocity $u^\mu$ and we have defined
\begin{align}
    Q_1(\beta)&=(7 + \beta^2) \gamma^2 E(\beta) - 4 K(\beta)\,,\\
    Q_2(\beta)&=(3 + 7 \beta^2 - 2 \beta^4) \gamma^2 E(\beta)- (3 + \beta^2) K(\beta)\,,\\
\nonumber   Q_3(\beta)&=(6 - 15 \beta^2 + 5 \beta^4 - 4 \beta^6) \gamma^2 E(\beta) \\
    &\hspace{80pt}- 2(3 - 6 \beta^2 + \beta^4) K(\beta)\,,
\end{align}
with $K(\beta)$ and $E(\beta)$ the complete elliptic integrals of the first and second kind [see Eq.~\eqref{eq:elliptic_def}]. We have also defined [see Eq.~\eqref{eq:picomps}]
\begin{align}
    \pi_{(0)} &= \pi^{xx} + \pi^{yy}\,,\\
 \pi_{(\pm 2)} &= \pi^{xx} \pm 2i \pi^{xy} - \pi^{yy}\,.
\end{align}
Note that by construction, Eq.~(\ref{eq:ep3}) can only be trusted close to equilibrium. In particular, it is not zero when $f$ is time-evolved according to free-streaming. We extract all final state observables at time $\tau=3R$, where their evolution has generally plateaued~\cite{Ambrus:2021fej,Ambrus:2022koq} due to the system cooling down and interaction rates dropping off.

\begin{figure}[t]
    \centering
    \includegraphics[width=.92\linewidth]{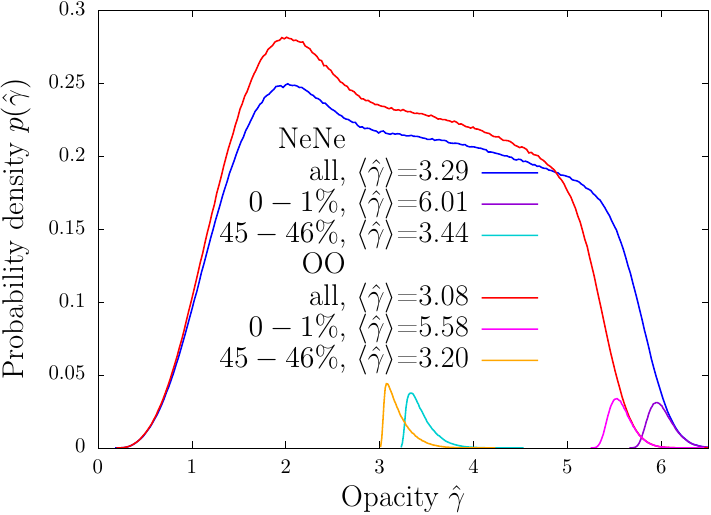}
    \caption{Distributions of event-by-event opacities $\hat{\gamma}$ for $\eta/s=0.12$ in the full ensembles of \trento Ne+Ne and O+O events for the parameters used in this work, as well as in the centrality classes 0-1\% and 45-46\%.}
    \label{fig:hist1d}
\end{figure}

\section{Characterizing initial states}

\label{sec:3}

We will now describe some of the properties of the initial state ensemble that are relevant for the phenomenology we are interested in. Even on an event-by-event level, the opacity $\hat{\gamma}$ largely determines the strength of flow responses to initial state properties~\cite{Ambrus:2024hks}. Fig.~\ref{fig:hist1d} shows for a shear viscosity of $\eta/s=0.12$ the distributions of event-by-event opacities for the full ensembles of Ne+Ne and O+O collisions, as well as of the centrality classes for which we simulated the collision dynamics. The full distributions are mostly flat with dropoffs at the low and high ends and cover a range of opacities between 1 and 6. As expected, the 0-1\% classes make up the high end of the distribution, while the 45-46\% classes are in the middle. The spread of opacities within each class is small, such that flow response coefficients can be expected to vary little within each class. However, between O+O and Ne+Ne opacities in the same centrality classes differ slightly, so that we can expect flow responses to show slight differences, which is discussed in Sec.~\ref{sec:flow_ratios} and will need to be taken into account when inferring initial state information from flow ratios. 

On top of this, the medium starts converting eccentricities in transverse position space into anisotropic flow when it starts expanding transversely. Any prior change to the eccentricities will bring an additional modulation of the ratio of initial state eccentricities to final state flow. Such a change can in fact be brought about via inhomogeneous cooling~\cite{Ambrus:2021fej}, i.e., the fact that hotter regions cool more than colder ones, which changes the weighting of different parts of the transverse plane and therefore the eccentricities. This effect depends more strongly on details of the geometry than the flow response, so it can manifest in different ways in O+O and Ne+Ne collisions and thereby cause a difference in how eccentricity ratios translate to flow ratios.

\begin{figure}[t]
    \centering
    \includegraphics[width=\linewidth]{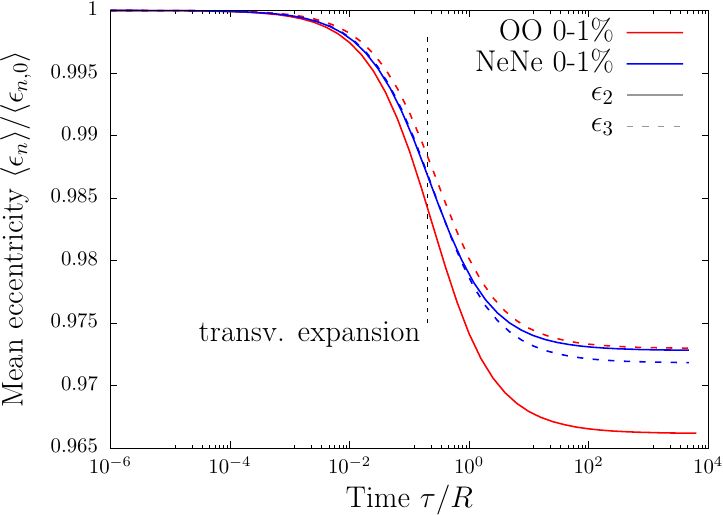}
    \caption{Time evolution of ensemble-averaged ellipticity $\langle\epsilon_2\rangle$ and triangularity $\langle\epsilon_3\rangle$ in local Bjorken flow, normalized to its initial value, for $\eta/s=0.12$. In the absence of transverse expansion, all change is due to inhomogeneous cooling. The vertical dashed line is at $\tau/R=0.2$, which would be the approximate timescale of the onset of transverse expansion in a 2+1D description~\cite{Ambrus:2022koq}.}
    \label{fig:cooling}
\end{figure}

\begin{figure}[t]
    \centering
    \includegraphics[width=.78\linewidth]{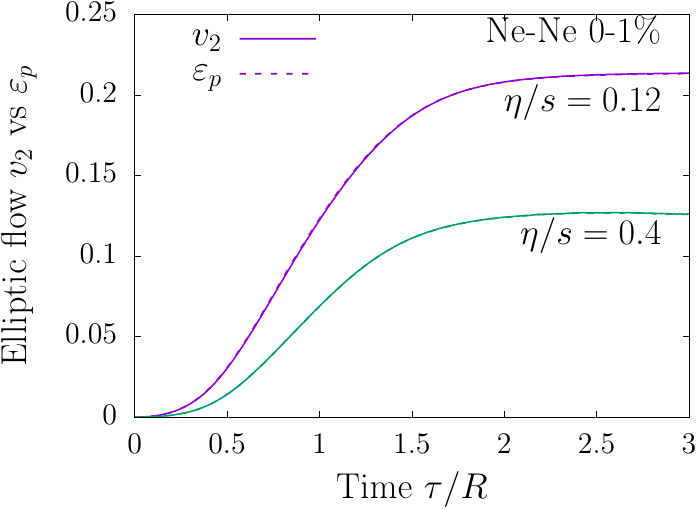}\\
     \includegraphics[width=.78\linewidth]{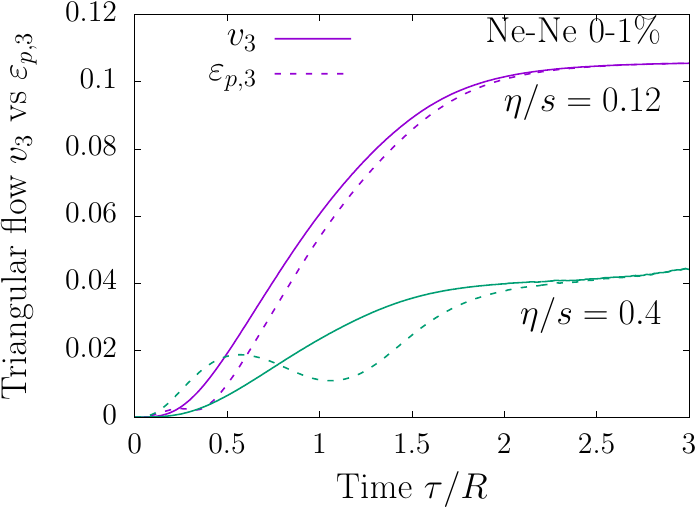}
    \caption{Time evolution of (top) elliptic flow in terms of $v_2$ (solid lines) and $\varepsilon_p$ (dashed lines) as well as (bottom) triangular flow in terms of $v_3$ (solid lines) and $\varepsilon_{p,3}$ (dashed lines) at two different values of the shear viscosity $\eta/s$ in different colors for an example event from the 0-1\% centrality class of Ne+Ne collisions. }
    \label{fig:epvsv}
\end{figure}

We check this effect on our initial state ensembles by propagating all events in local Bjorken flow~\cite{Ambrus:2021fej}, a model description that assumes that each point in the transverse plane evolves as if it were a transversely homogeneous system. This is an accurate description of the actual evolution up until transverse expansion sets in. At a shear viscosity of $\eta/s=0.12$, Fig.~\ref{fig:cooling} shows the time evolution of average $\epsilon_2$ and $\epsilon_3$ in the 0-1\% centrality classes, for O+O and Ne+Ne collisions. At the final time, the change in $\epsilon_3$ is similar in both systems, but, with $3.4\%$, the ellipticity in O+O collisions decreases more than in Ne+Ne collisions, with $2.5\%$. However, around the timescale of the onset of transverse expansion, the eccentricities in both systems have decreased by a comparable fraction of around $1.5\%$, so we do not expect this effect to manifest in neon-over-oxygen ratios of final-state flow coefficients.

 \begin{figure*}[t]
     \centering
    \includegraphics[width=.46\linewidth]{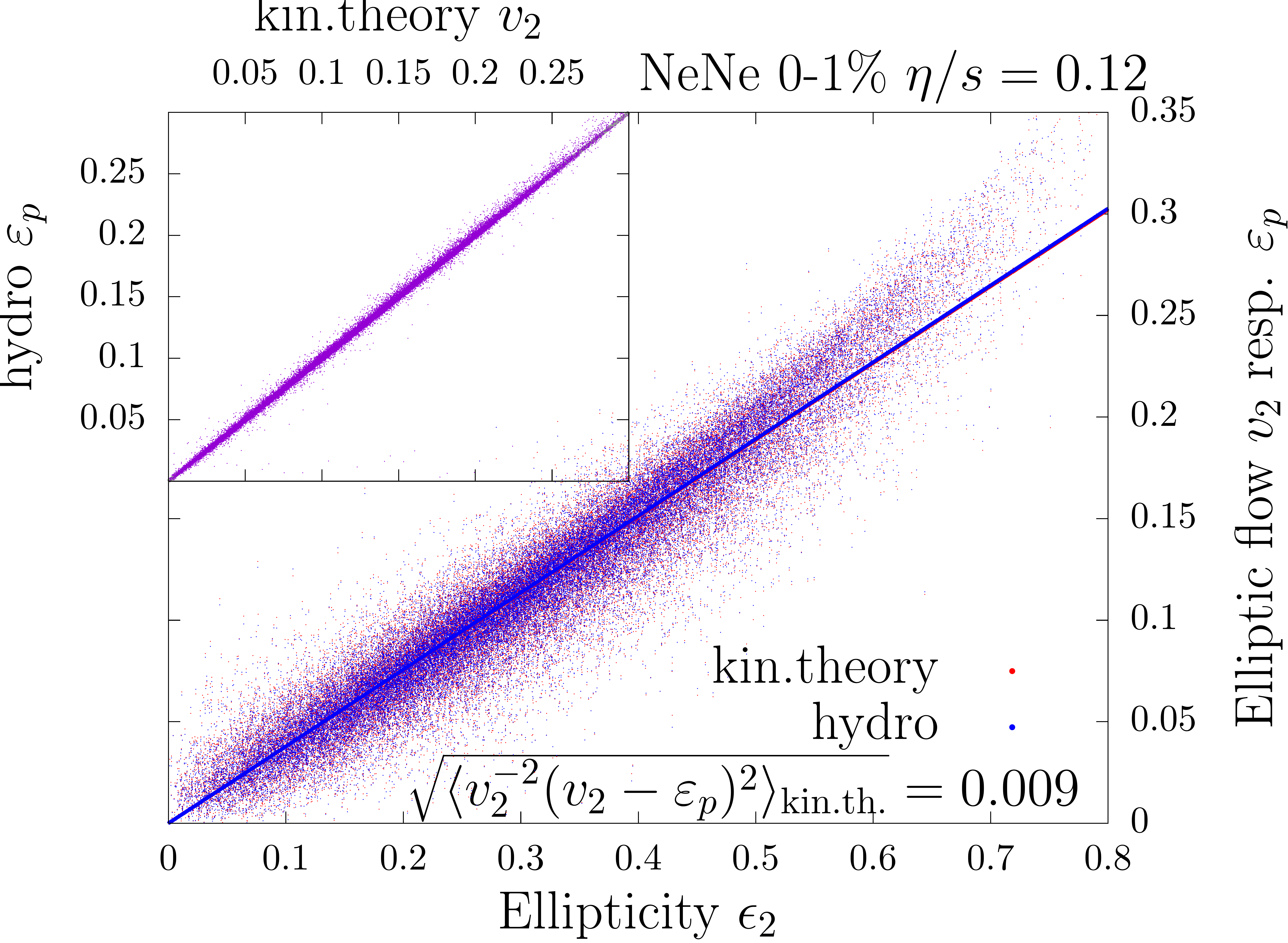}\hspace{20pt}
     \includegraphics[width=.46\linewidth]{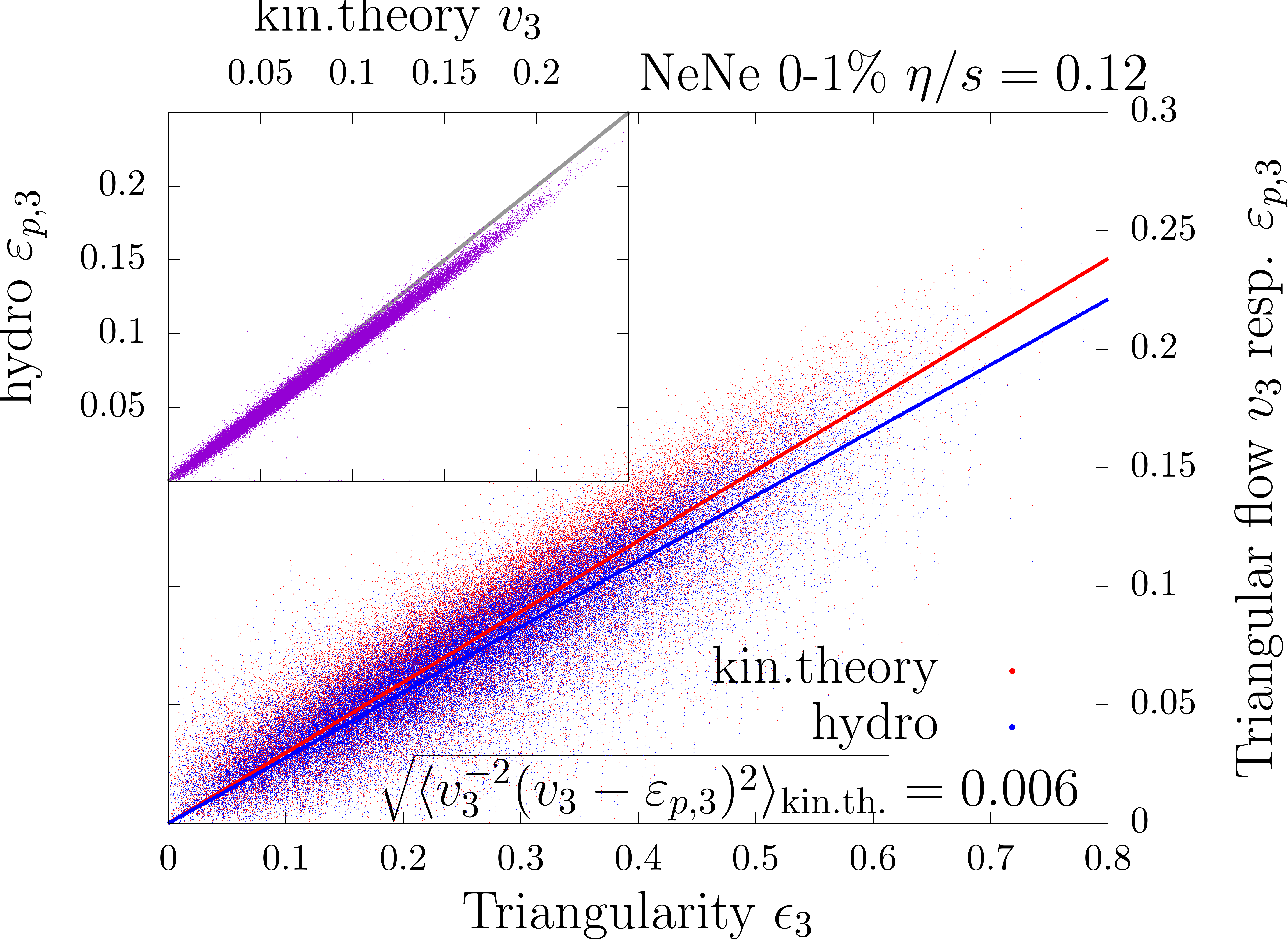}
     \caption{Event-by-event results for (left) elliptic flow versus ellipticity and (right) triangular flow versus triangularity from kinetic theory as $v_n$ (red) and hydrodynamics as $\varepsilon_{p(,3)}$ (blue) from the 0-1\% centrality class of Ne+Ne collisions at $\eta/s=0.12$. Lines in the main panels are linear fits to the scatter plots.}
     \label{fig:ebeepvsv}
 \end{figure*}

\section{Results}
\label{sec:4}

\subsection{Event-by-event $v_2$, $v_3$, and their proxies}

We start by analyzing event-by-event flow responses, to get an idea of whether our proxy measures of $v_2$ and $v_3$ in hydrodynamics, respectively, $\varepsilon_p$ and $\varepsilon_{p,3}$, correctly track the flow coefficients of the kinetic theory.

In fact, a direct comparison of these observables can be performed in kinetic theory, where all are accessible in terms of the phase space distribution, $f$. Figure~\ref{fig:epvsv} compares the time evolution of the hydrodynamic proxies with the observables they are meant to represent for an example event. For elliptic flow, the proxy is a minor reweighting by $p_\perp/p^0$. In this case, no difference is visible between the two definitions. The proxy for triangular flow is defined based on a linearization around equilibrium, which thus degrades for larger $\eta/s$, and is not a good assumption at early times, as the plot also shows. However, by the time we extract the final state flow at $\tau=3R$, the two have come into good agreement.

For completeness we also look at event-by-event flow responses in the 0-1\% centrality class of Ne+Ne at $\eta/s=0.12$. Appendix~\ref{app:spread} contains similar plots for the 45-46\% centrality class and $\eta/s=0.4$. The main plots in Fig.~\ref{fig:ebeepvsv} show point clouds of final-state flows plotted against initial-state eccentricities, for both kinetic theory and hydrodynamics, together with straight lines indicating the average flow response coefficient. As in earlier work~\cite{Ambrus:2024hks}, the flow response from hydrodynamics falls short of that from kinetic theory, as the dynamics are not reproduced exactly. However, in contrast to that earlier work, there now seems to be a larger spread of event-by-event flow responses. As will be elaborated on a bit more in Sec.~\ref{sec:flow_responses}, this is due to the larger effect of fluctuations on the initial state geometry. Ref.~\cite{Ambrus:2024hks} considered nucleon widths of $w\approx1\,$fm with no substructure, and hence very smooth initial-state geometries. Now, our nucleons have a width of $w\approx0.6\,$fm, and are made of 3 constituents with a width of $v=0.5\,$fm, leading to larger local density fluctuations and eccentricities.

The bottom right labels in Fig.~\ref{fig:ebeepvsv} show that differences between the final state $v_n$ and $\varepsilon_{p(,3)}$ within kinetic theory are tiny also on average, so the proxy flow observables work as intended. The inset plots in he top left show the event-by-event comparison of the hydrodynamic proxies to the kinetic theory results for elliptic and triangular flow. In the case of elliptic flow, it is a very thin straight point cloud with slope almost 1. For triangular flow, the spread is a bit larger, but still there is a strong correlation. The slope deviates from 1, as the hydro response is weaker. Overall, though, also when taken from hydrodynamic simulations, the proxy flow observables measure what they are intended to measure.  % This is even more true when considering averages over centrality classes, in which a symmetric spread will cancel. 

\subsection{Correlations with the orientation angle}\label{sec:thetacorr}

\begin{figure*}[t]
    \centering
    \includegraphics[width=.32\linewidth]{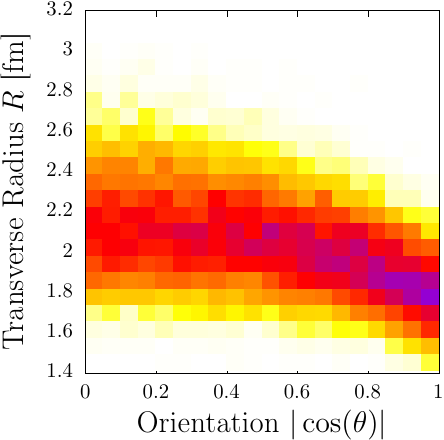}~
    \includegraphics[width=0.32\linewidth]{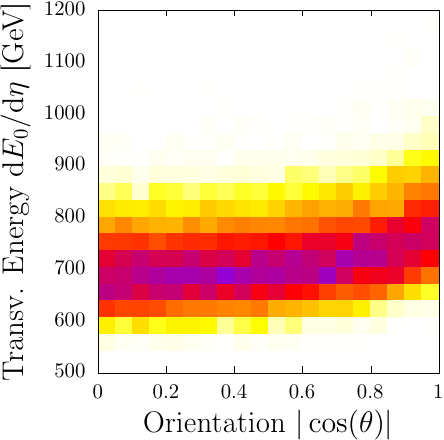}~
    \includegraphics[width=.32\linewidth]{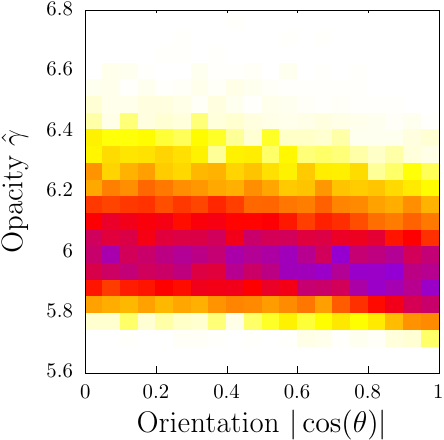}\\ \vspace{5pt}
    \includegraphics[width=0.32\linewidth]{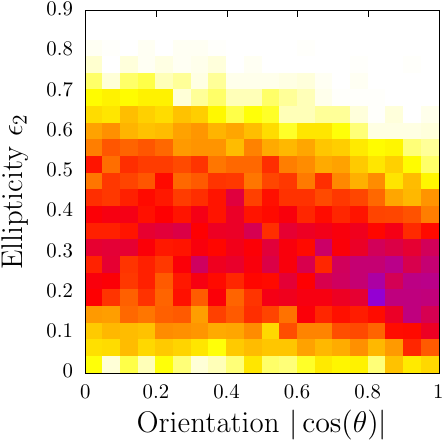}~
    \includegraphics[width=.32\linewidth]{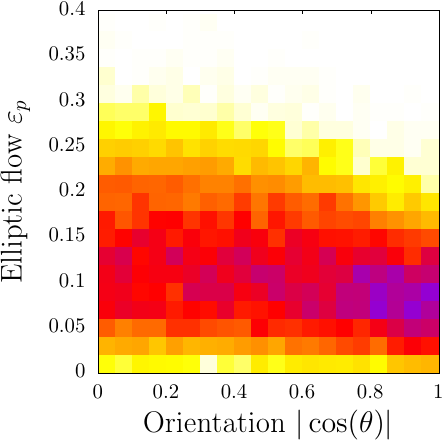}~
    \includegraphics[width=.32\linewidth]{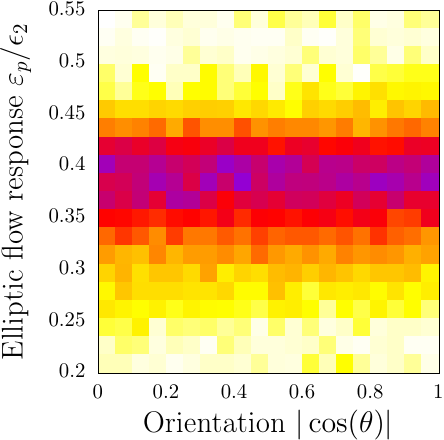}
    \caption{2D histograms of the joint distribution of Ne+Ne events from the 0-1\% centrality class in the orientation angle $\theta$ via $|\cos\theta|$ and quantities related to the flow response. Flow responses were simulated in kinetic theory at $\eta/s=0.12$. On both axes, the events were grouped into 20 bins.}
    \label{fig:thetacorr}
\end{figure*}

In collisions of nuclei with a prolate deformation, the orientation of the tip of these nuclei to the beam axis affects the initial properties of the created fireball. When considering very central collisions, it can be expected that the two nuclei have almost perfect overlap. In a collision where both tips point in the direction of the beam, the overlap area is smaller and less elliptic than in a collision where both tips point perpendicular to it. This leads in particular to an anti-correlation between elliptic flow and the mean transverse momentum \cite{Giacalone:2019pca,Giacalone:2020awm}, demonstrated experimentally in U+U collisions~\cite{STAR:2024wgy}. Now if the flow response also depends on this orientation angle to the beam axis, these initial state correlations will be modulated before being imprinted on the final state, which changes how the measured data have to be interpreted. 

Within the PGCM setup, we can check in our simulations to what extent this is true. For this, we have divided the orientation angle into 20 bins in $|\cos\theta|$ and selected 19\,340 events from the 0-1\% centrality class of Ne+Ne where the orientation angles of both nuclei fall into the same bin. We selected the same amount of events for each bin. Fig.~\ref{fig:thetacorr} shows heatmaps of the correlations of several observables related to elliptic flow with $|\cos\theta|$. The observables were also divided into 20 bins. As stated earlier, the opacity mostly determines the strength of the flow response. The first row shows correlations of the opacity $\hat{\gamma}$ and the two initial state properties it depends on: the rms transverse radius $R$ and the transverse energy $\d E_\perp^0/\d\eta$. As expected from the discussion above, $R$ decreases when the nuclei are more aligned with the beam axis. At the same time, the transverse energy increases, likely because there is a larger number of collisions of nucleon pairs. As a result, the distribution of opacities depends only marginally on the orientation angle.

The second row shows how correlations in the ellipticity $\epsilon_2$ translate to correlations in elliptic flow $\varepsilon_p$ and the response $\varepsilon_{p}/\epsilon_2$. Again as described earlier, ellipticities are smaller if the nuclei are more aligned with the beam axis, which in this heatmap plot results in a narrower distribution with smaller mean at larger $|\cos\theta|$. The elliptic flow responses replicate the same pattern, so it seems that these correlations are transported faithfully to the final state through the flow responses. Indeed, the distribution of flow responses $\epsilon_p/\epsilon_2$ shows no discernible dependence on $|\cos\theta|$.

\subsection{Flow responses versus opacity}
\label{sec:flow_responses}

\begin{figure*}
    \centering
    \includegraphics[width=0.46\linewidth]{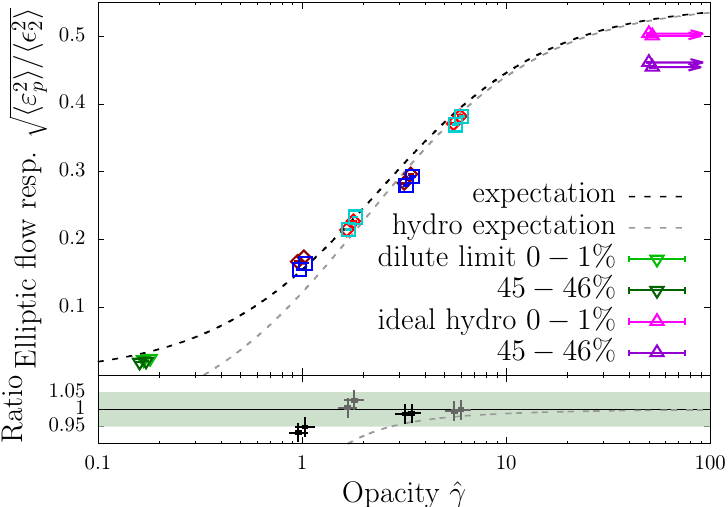}\hspace{20pt}
    \includegraphics[width=0.46\linewidth]{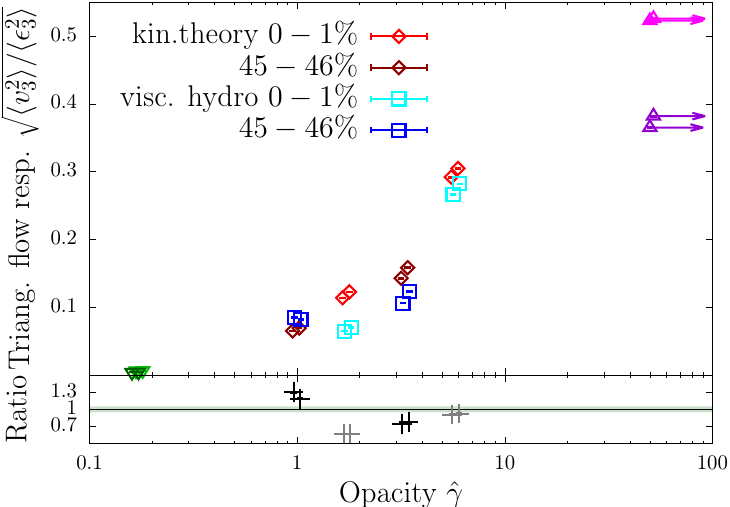}
    \caption{(Left) elliptic and (right) triangular flow responses in kinetic theory and hydrodynamic simulations, estimated through the ratio of rms flow $\sqrt{\langle\varepsilon_p^2\rangle}$ resp. $\sqrt{\langle v_3^2\rangle}$ to rms eccentricities $\sqrt{\langle\epsilon_n^2\rangle}$, plotted against the mean opacity $\langle\hat{\gamma}\rangle$ and compared to the expectation from previous results~\cite{Ambrus:2024hks}. The bottom part of the plot shows the ratio of hydrodynamic to kinetic theory results with a green band indicating the 5\% accuracy level. Different points are obtained from two different centrality classes and paired in O+O (lower $\hat \gamma$) and Ne+Ne (higher $\hat \gamma$) collisions, and for four cases of the shear viscosity: $\eta/s=0.12$ (higher $\hat{\gamma}$) and $\eta/s=0.4$ (lower $\hat{\gamma}$) as well as the ideal ($\eta/s=0$) and dilute ($\eta/s=\infty$) limits.}
    \label{fig:kappagamma}
\end{figure*}

Once eccentricities are known, the main factor determining the strength of flow responses is the opacity. In fact, in earlier works based on the same simulation codes for RTA kinetic theory and hydrodynamics, mean flow responses $\kappa_{2,2}=\varepsilon_p/\epsilon_2$ followed a universal flow response curve $\kappa_{2,2}(\hat{\gamma})$~\cite{Ambrus:2024hks,Ambrus:2024eqa}. Similar results have been found in kinetic theory with a $2\leftrightarrow 2$ collision kernel in the conformal~\cite{Nugara:2024net} and non-conformal case~\cite{Nugara:2025ueb}.  In Fig.~\ref{fig:kappagamma}, we plot kinetic theory and hydrodynamic flow responses in the 0-1\% and 45-46\% centrality classes of O+O and Ne+Ne collisions through the ratio of the rms elliptic (left panel) and triangular (right panel) flows to the corresponding rms eccentricities against the mean opacity. In the case of elliptic flow, we also plot (dashed lines) the universal flow response curves obtained in Refs.~\cite{Ambrus:2024hks,Ambrus:2024eqa}.

 Comparing hydrodynamics to kinetic theory, the elliptic flow responses appear in agreement for a wider range in opacity than in the previous works, as the 5\% accuracy level has moved from $\hat{\gamma} \approx4$ down to $\hat{\gamma}\approx1$. This is likely a coincidence, and due to higher order responses deviating in a different direction than the linear one, as will be corroborated in the following. Triangular flow responses decrease in strength faster with decreasing opacity, with flow responses at $\hat{\gamma}=3$ only about a third as strong as in the ideal case. Similarly, differences between kinetic theory and hydrodynamics appear already at higher opacities and are on the order of 10\% for the 0-1\% centrality
classes with realistic values of the shear viscosity ($\eta/s=0.12$), which has a mean opacity of $\langle\hat{\gamma}\rangle\approx6$. This confirms the well-known feature that higher order harmonic flow is more sensitive to viscous effects~\cite{Schenke:2010rr,Staig:2011wj,Teaney:2012ke,Plumari:2015cfa}, and therefore also more sensitive to the deviation from hydrodynamic behavior.

 One point worth mentioning is that, compared to our earlier work, while we observed in these new results a smooth transition of the flow response from an opacity-linear limit to a saturation in the ideal hydrodynamic limit at large opacities, the flow responses in different centrality classes no longer collapse onto the same curve. Differences between flow responses in the 0-1\% and 45-46\% centrality classes are on the order of 10\% for elliptic flow and 30\% for triangular flow. It then seems that these flow responses are not universal in fact.

\begin{figure*}[t]
    \centering
    \includegraphics[width=0.46\linewidth]{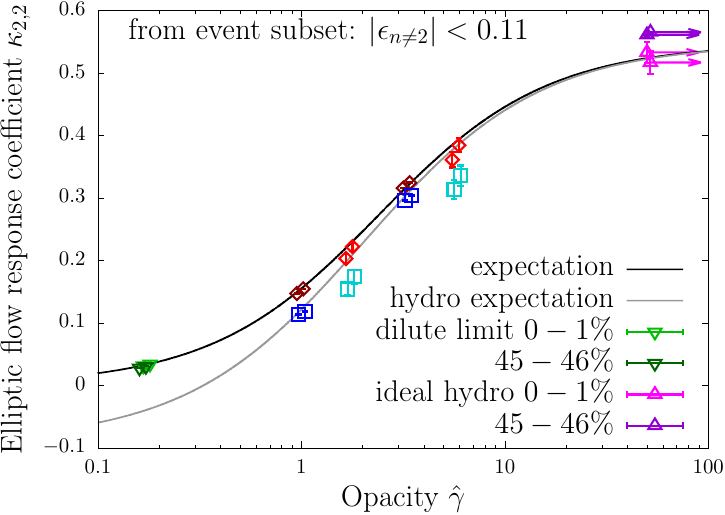}\hspace{20pt}
    \includegraphics[width=0.46\linewidth]{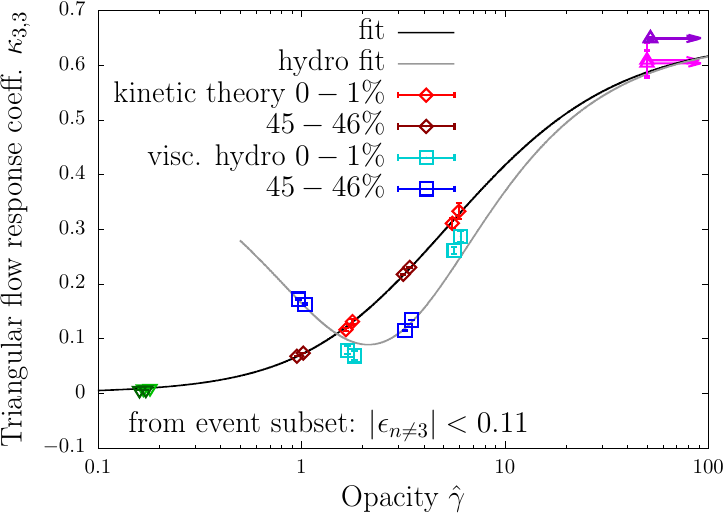}
    \caption{Linear order flow response coefficients $\kappa_{n,n}=v_n/\epsilon_n+\mathcal{O}(\epsilon_n)$ as a function of opacity $\hat{\gamma}$ for (left) elliptic flow, again compared to the expectation from the elliptic flow response curve obtained in Ref.~\cite{Ambrus:2024hks}, and (right) triangular flow, compared to the fitting functions in Eqs.~(\ref{eq:kappa33kinth}) and~(\ref{eq:kappa33hyd}). Different symbols are as in Fig.~\ref{fig:kappagamma}.}
    \label{fig:kappagammalin}
\end{figure*}

To investigate this further, in Fig.~\ref{fig:kappagammalin} we show the actual linear order flow response coefficients, dubbed  $\kappa_{n,n}$, and find that they fall on top of the previously obtained universal elliptic flow response curves. These coefficients are obtained through fits of the form
\begin{equation}
v_n=\kappa_{n;n}\epsilon_n+\kappa_{n;n,n,n}\epsilon_n^3   ,
\end{equation}
motivated by the findings of Refs.~\cite{Niemi:2012aj,Noronha-Hostler:2015dbi}. To suppress in particular the effects of non-linear couplings of harmonics, these coefficients are obtained from a fit of the event-by-event flow responses in the subset of events where all other eccentricities were small, $|\epsilon_m|<0.11$ for $m\neq n$. Note that this selection criterion reduces the statistics and, therefore, increases errors.

In the right panel, triangular flow responses can also be described by a single curve, so that we can meaningfully extract universal flow response curves through fits to the simulation points. The elliptic flow response curves were obtained in Ref.~\cite{Ambrus:2024hks} by fitting the Padé approximants
\begin{align}
    \kappa_{2,2}(\hat{\gamma})&=\frac{0.201\hat{\gamma}+0.129\hat{\gamma}^2}{1+0.892\hat{\gamma}+0.235\hat{\gamma}^2} ~~~[{\rm kinetic~theory}] \, ,\\
\kappa_{2,2}(\hat{\gamma})&=\frac{-0.0896+0.271\hat{\gamma}}{1+0.496\hat{\gamma}}~~~[{\rm hydrodynamics}],
\end{align}
where the large opacity limit of both curves was fixed through the ideal hydrodynamic results and the linear order limit of the kinetic theory curve through results from the opacity-linearized code LOebe. Analogously, we fit the triangular flow responses through (we retain values of $\langle \hat \gamma \rangle$ from the full sample before imposing the $|\epsilon_m|$ constraints)
\begin{align}
\kappa_{3,3}(\hat \gamma)&=\frac{0.0441\hat{\gamma}+0.0964\hat{\gamma}^2}{1+0.833\hat{\gamma}+0.149\hat{\gamma}^2}\label{eq:kappa33kinth} ~~~[{\rm kinetic~theory}], \\
    \kappa_{3,3}(\hat \gamma)&=\frac{0.464-0.330\hat{\gamma}+0.0955\hat{\gamma}^2}{1+0.249\hat{\gamma}+0.147\hat{\gamma}^2}~~[{\rm hydrodynamics}].
    \label{eq:kappa33hyd}
\end{align}
The fitted curves are shown in Fig.~\ref{fig:kappagammalin}. In comparison to the elliptic flow curve, the fitting function to the hydrodynamic triangular flow has one more free parameter, which is necessary to capture the non-monotonic behavior. Note that this non-monotonic behavior is not due to any difference between taking $\varepsilon_{p,3}$ in hydrodynamics and $v_3$ in kinetic theory. From the kinetic simulations, these quantities are actually found to be near identical, as seen in Fig.~\ref{fig:ebeepvsv} and Figs.~\ref{fig:ebeepvsvl}--\ref{fig:ebeepvsv45l}. Therefore, this behavior represents one of our main results. Especially for peripheral collisions, the triangular flow response is strongly impacted by nonlinear terms, leading to qualitatively different trends between hydrodynamics and kinetic theory.

Likely, the reason why the centrality-dependent flow responses in Fig.~\ref{fig:kappagamma} do not follow universal curves is due to contributions from nonlinear responses~\cite{Gardim:2011xv,Teaney:2012ke,Kirchner:2026hpw}, which are stronger here than in our previous work because of the larger initial-state fluctuations driven by smaller nucleon sizes, leading to higher eccentricities for all $n$. These nonlinear couplings will have unique dependencies on the opacity. For completeness, the full flow estimator can take the following form in terms of complex variables $\mathcal{E}_n=\epsilon_ne^{in\Psi_n}$ and $\mathcal{V}_n=v_ne^{in\Psi_{p,n}}$:
\begin{align}
         \mathcal{V}_n \approx &\kappa_{n;n}(\hat{\gamma})\mathcal{E}_n\nonumber\\
         &+\sum_{1\le m\le n/2} \kappa_{n;m,|n-m|}(\hat{\gamma})\mathcal{E}_m\mathcal{E}_{|n-m|}\nonumber\\
         &+\sum_{m\ge 1}\kappa_{n;m,n+m}(\hat{\gamma})\mathcal{E}_m^*\mathcal{E}_{n+m}\label{eq:full_flow}\\
         &+\sum_{\substack{1\le m\le n/2+1\\1\le l\le (n-m)/2}} \kappa_{n;m,l,|n-m-l|}(\hat{\gamma})\mathcal{E}_m\mathcal{E}_l\mathcal{E}_{|n-m-l|}\nonumber\\
         &+\sum_{\substack{m\ge 1\\1\le l\le (n+m)/2}} \kappa_{n;m,l,|n+m-l|}(\hat{\gamma})\mathcal{E}_m^*\mathcal{E}_l\mathcal{E}_{|n+m-l|}\nonumber\\
         &+\sum_{m,l\ge 1} \kappa_{n;m,l,n+m+l}(\hat{\gamma})\mathcal{E}_m^*\mathcal{E}_l^*\mathcal{E}_{n+m+l}\nonumber\\
         &+\mathcal{O}(\mathcal{E}_m^4)\nonumber\,,
    \end{align}
    where all terms ensure $n$-fold rotational symmetry. We have neglected moments with higher order powers of the radius~\cite{Teaney:2010vd}, i.e., highe-order radial eccentricities, as their influence is likely subleading. A determination of the nonlinear flow response curves and examination of Eq.~\eqref{eq:full_flow} is left for future work.

\subsection{Flow ratios: nuclear structure and nonequilibrium dynamics}\label{sec:flow_ratios}

\begin{figure*}
    \centering
    \includegraphics[width=0.46\linewidth]{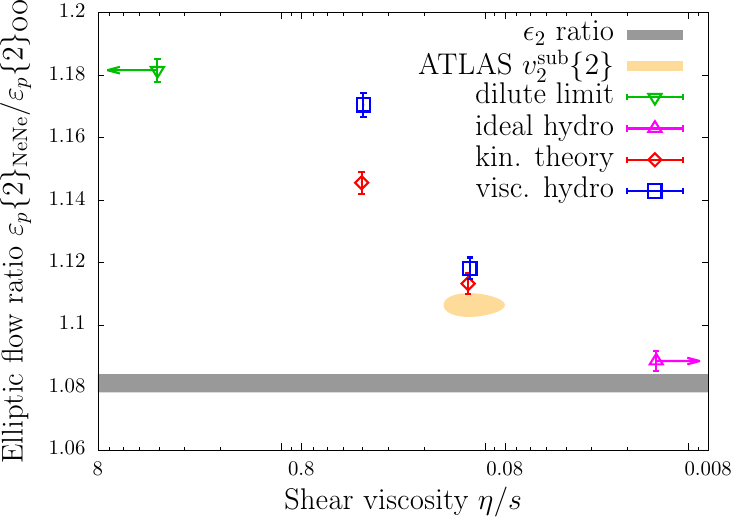}\hspace{20pt} \includegraphics[width=0.46\linewidth]{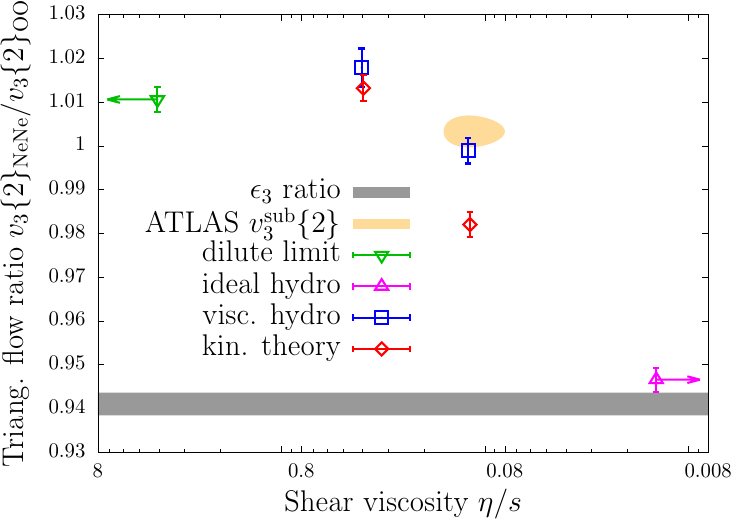}
    \caption{Flow ratios between the 0-1\% centrality classes of Ne+Ne and O+O collisions for (left) elliptic and (right) triangular flow extracted from kinetic theory and hydrodynamic simulation results at four different values of the shear viscosity: $\eta/s=0.12$ and $\eta/s=0.4$ as well as the ideal hydrodynamic ($\eta/s=0$) and dilute ($\eta/s=\infty$) limits. They are compared to the corresponding eccentricity ratios in the initial state as well as ATLAS measurements taken from Ref.~\cite{ATLAS:2025nnt}, where the horizontal error is an estimate from Bayesian fits of $\eta/s$, which together with the experimental error is plotted as a one-$\sigma$ ellipse.}
    \label{fig:flow_ratio}
\end{figure*}

We now assess the robustness of the signals of the deformation of in particular $^{20}$Ne on the flow results, trying to establish whether a hydrodynamic framework would be appropriate for the inference of correlation properties of light ions from O+O and Ne+Ne data. For this purpose, we focus first on ultracentral collisions with near-full nuclear overlap, which are the most impacted by nuclear shapes, and consider ratios of flow coefficients taken between the two collision systems, as they largely cancel common initial-state features, flow responses, and associated uncertainties \cite{Giacalone:2024luz}. Ratios of flow coefficients were in particular measured by the LHC collaborations, whose results are consistent with each other~\cite{ATLAS:2025nnt,ALICE:2025luc,CMS:2025tga}. Although our setup is not meant to be compared directly to measurements,  it has been demonstrated~\cite{Kurkela:2019kip} that the main effect of our assumptions is a rescaling factor on flow coefficients, such that meaningful theory-to-data comparisons should be possible for ratios.

Our ratio results for rms ultracentral anisotropies, $v_n\{2\}\equiv\sqrt{\langle v_n^2 \rangle}$, and analogously for $\varepsilon_{p}$, are shown in Fig.~\ref{fig:flow_ratio} as a function of the coupling, here $\eta/s$. The left panel shows elliptic flow.  The first remark is that the ratio tends to increase from the ideal-hydrodynamic to the dilute-transport limit. This means that, as viscous corrections increase, the flow of O+O collisions decreases more quickly than that of Ne+Ne collisions. For a value of $\eta/s=0.12$, agreeing with Bayesian extractions \cite{Nijs:2025qxm,Heffernan:2023gye}, our viscous hydro and kinetic results are in good agreement with the ATLAS measurement, and are indistinguishable from each other. Significant deviations among the two models occur instead for $\eta/s=0.4$. Our conclusion is thus that the impact of the deformation of $^{20}$Ne on elliptic flow can be assessed from the ratio data with little uncertainty on the dynamical modeling. 

A similar pattern emerges in the right panel of Fig.~\ref{fig:flow_ratio}. At the phenomenologically relevant $\eta/s$, however, the triangular-flow ratios predicted by kinetic theory and hydrodynamics differ by about 2\%, a significant difference given the precision of the experimental measurements. The octupole (or tetrahedral) deformation of oxygen is expected to modify the rms triangular flow by approximately 10\%~\cite{YuanyuanWang:2024sgp}, comparable to the difference in absolute triangular-flow responses between the two frameworks shown in Fig.~\ref{fig:kappagamma}. The smaller dynamical dependence of the neon-to-oxygen ratio thus makes it a promising observable for isolating nuclear-structure effects. Nevertheless, the impact of deformation and clustering on this ratio is less well-understood than on the individual O+O value. This residual uncertainty should therefore be taken into account when interpreting $v_3$ ratios.

\begin{figure*}[t]
    \centering
    \includegraphics[width=0.46\linewidth]{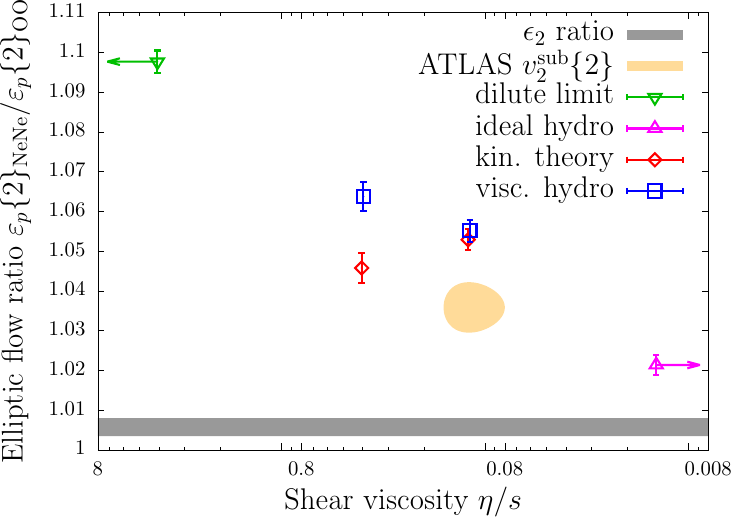} \hspace{20pt} \includegraphics[width=0.46\linewidth]{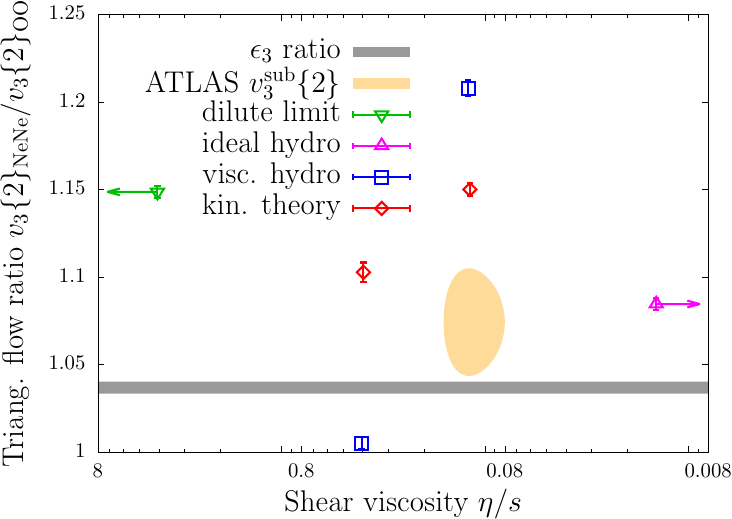}
    \caption{Same as Fig.~\ref{fig:flow_ratio}, but for the 45-46\% centrality classes}
    \label{fig:flow_ratio_m}
\end{figure*}

To conclude, we show ratios in the 45-46\% centrality class in Fig.~\ref{fig:flow_ratio_m}. In this class, effects of nuclear shapes should be essentially absent. For elliptic flow in the left panel, the trends are qualitatively similar to those shown in Fig.~\ref{fig:flow_ratio}, though the simulations do not capture the ATLAS data as well. Conversely, the triangular flow results in the right panel show some dramatic behavior, due to the steeper variations with centrality and opacity of the response. First, we see that the trend of the ratio is no longer monotonic as a function of $\eta/s$. More important, for $\eta/s=0.12$, the difference between kinetic theory and hydrodynamics is larger than 5\%. As also reported in the experimental papers \cite{ALICE:2025luc,CMS:2025tga}, hydro results tend to overshoot the experimental data. Remarkably, then, switching to kinetic theory largely improves agreement. This points to the impact of out-of-equilibrium effects that are not captured by hydrodynamics, and that may be necessary to reproduce the experimental values. Given that ratios reduce uncertainties related to the modeling of the initial states, experimental flow ratios of light ions in peripheral collisions then appear to have high potential as probes of genuine nonequilibrium phenomena, which deserves further investigation.

\section{Conclusion \& Outlook}
\label{sec:5}

Light-ion collisions offer a way to investigate the dynamics of small systems
with initial-state geometries constrained by microscopic nuclear theory.
At the same time, their collective flow reveals information about
correlations in the colliding nuclear quantum many-body states.
In this work, we assessed these complementary opportunities through
event-by-event hydrodynamic and conformal RTA kinetic-theory calculations of
O+O and Ne+Ne collisions, using nuclear configurations
informed by \textit{ab initio} PGCM calculations. 

Compared with our previous studies, the initial conditions used here,
informed by more recent Bayesian analyses, contain stronger local fluctuations
and larger eccentricities. The resulting effective flow responses, defined
through ratios of rms flow to rms eccentricity, no longer follow a common
curve determined solely by the opacity. Approximate universality is recovered
for the purely linear response coefficients. This points to
an enhanced role for nonlinear response, including couplings between
harmonics, in shaping the final flow. Determining the opacity dependence of
these additional responses is a natural extension of the present
analysis.

For nuclear-structure studies, the most favorable regime is that of
ultracentral collisions. At $\eta/s=0.12$, the Ne+Ne-to-O+O elliptic-flow
ratio agrees at the subpercent level between hydrodynamics and kinetic
theory, while the triangular-flow ratio differs by approximately $2\%$.
This cancellation is significant, as the absolute triangular-flow responses
differ by about $10\%$ even in this centrality class. We also find little
dependence of the ultracentral elliptic response on the orientation of
$^{20}$Ne, while the early modification of eccentricities by inhomogeneous
cooling is similar in the two collision systems. Together, these results
support the use of viscous hydrodynamics in Bayesian analyses aimed at
precisely constraining nuclear correlations in $^{20}$Ne and $^{16}$O from
ultracentral flow ratios. The residual dynamical dependence, particularly
for triangular flow, should be propagated as a modest theoretical uncertainty.

In regard to the applicability of hydrodynamics, peripheral collisions provide a more sensitive test of the dynamical
description. In the $45$--$46\%$ centrality class, triangular-flow ratios
exhibit a substantially larger separation between hydrodynamics and kinetic
theory than in ultracentral collisions. At $\eta/s=0.12$, the
kinetic-theory prediction lies closer to the measured ratio. The enhanced
sensitivity of triangular flow to the evolution beyond hydrodynamics makes
peripheral light-ion flow ratios promising probes of nonequilibrium 
dynamics. Establishing their full discriminating power will require a
systematic treatment of the remaining initial-state uncertainties and of
the relation between the energy-weighted observables studied here and the
measured hadronic flow.

As an outlook, two extensions are particularly promising. First, higher harmonics such
as quadrangular flow contain substantial nonlinear
contributions~\cite{Kurkela:2020wwb,Ambrus:2022koq} and may
therefore provide even greater sensitivity to departures from hydrodynamic
behavior. This expectation can be tested by extending the present comparison
to $v_4$ and its nonlinear components. Preliminary ALICE measurements of
$v_4\{2\}$ in both O+O and Ne+Ne collisions already show encouraging
precision~\cite{Machacek:2026ilw}. Second, symmetric collisions of still lighter nuclei would provide access
to lower opacities while retaining initial-state geometries constrained by
\textit{ab initio} nuclear structure. A particularly compelling case is
$^{4}$He collisions, recently explored in kinetic theory calculations in~\cite{Taghavi:2026vyw}. Even a short
dedicated pilot run during LHC Run~4, as considered in recent accelerator
studies~\cite{NewIonCollisionsLHCRun4}, could thus open a new experimental discovery
window for out-of-equilibrium QCD.

\begin{acknowledgements}
We thank the participants of the ExtreMe Matter Institute EMMI workshop on Light Ion Collisions 2026 for discussions related to the topic of this work.
    CW has received funding from the European Research Council (ERC) under the European Union’s Horizon 2020 research and innovation programme (grant number: 101089093 / project acronym: High-TheQ). Views and opinions expressed are however those of the authors only and do not necessarily reflect those of the European Union or the European Research Council. Neither the European Union nor the granting authority can be held responsible for them. FZ acknowledges support by the DFG through Emmy Noether Programme (project number 496831614) and CRC 1225 ISOQUANT (project number 27381115) as well as the state of Baden-Württemberg for its support of bwHPC and the German Research Foundation (DFG) for funding under "Project number 455622343" (bwForCluster NEMO 2). V.E.A. gratefully acknowledges the support of the Alexander von Humboldt Foundation through a Renewed Research Stay Fellowship, as well as support by the Deutsche Forschungsgemeinschaft (DFG, German Research Foundation) through the CRC-TR 211 ``Strong-interaction matter under extreme conditions,'' Project No. 315477589-TRR 211. Numerical calculations presented in this work were performed at the Paderborn Center for Parallel Computing (PC2) and the Center for Scientific Computing (CSC) at the Goethe-University of Frankfurt.
\end{acknowledgements}

\appendix

\section{Milne coordinates}
\label{Milne_coordinates}

In this Appendix, we present in more detail how to rewrite the Boltzmann equation~\eqref{eq:boltz} in Milne coordinates suitable for describing a longitudinally expanding system, see Eq.~\eqref{eq:Milne_coordinates}. The inverse transformations are given by 
\begin{equation}
    t = \tau \cosh \eta, \quad z = \tau \sinh\eta.
\end{equation}
The Minkowski line element is modified to
\begin{equation}
 ds^2 = dt^2 - d\mathbf{x}_\perp^2 - dz^2 = d\tau^2 - d\mathbf{x}_\perp^2 - \tau^2 d\eta^2,
\end{equation}
such that $\sqrt{-g} = \tau$. The contravariant components of the particle momentum expressed with respect to the Bjorken coordinates read $p^\mu = (p^\tau, \mathbf{p}_\perp, p^\eta)$, where 
\begin{equation}
 p^\tau = \frac{t}{\tau} p^t - \frac{z}{\tau} p^z, \quad 
 p^\eta = -\frac{z}{\tau^2} p^t + \frac{t}{\tau^2} p^z.
\end{equation}
From the momentum rapidity given by $\tanh y = p^z/p^t$ or inversely,
\begin{equation}
    p^t = p_\perp \cosh y, \quad p^z = p_\perp \sinh y,
\end{equation}
it is easy to see that 
\begin{equation}
 \tanh(y - \eta) = \frac{\tau p^\eta}{p^\tau},
\end{equation}
or equivalently, $p^\tau = p_\perp \cosh(y - \eta)$ and $\tau p^\eta = p_\perp \sinh(y - \eta)$.
The generally covariant integration measure becomes 
\begin{equation}
 dP = \nu_{\rm eff} \frac{d^2\mathbf{p}_\perp}{(2\pi)^3} d(y-\eta).
\end{equation}
From this, one straightforwardly obtains the initial energy density in Eq.~\eqref{eq:init_e_dens}.

\section{Numerical solution of the Boltzmann Equation}
\label{numerical_solution}

The total energy per unit rapidity $dE_0/d\eta$ as well as the system size $R$ allow us to introduce a reference energy density, a reference temperature and a reference length via
\begin{equation}
 \epsilon_{\rm ref} = \frac{dE_0 / d\eta}{\pi R^2 \tau_0}, \quad T_{\rm ref} = \left(\frac{\epsilon_{\rm ref}}{a}\right)^{1/4}, \quad L_{\rm ref} = (\tau_0 R^3)^{1/4}.
 \label{eq:ref}
\end{equation}
Then, the relaxation time $\tau_R$ can be written in terms of opacity $\hat{\gamma}$,
\begin{equation}
 \frac{1}{\tau_R} = \frac{\hat{\gamma} T}{T_{\rm ref} L_{\rm ref}}, \label{eq:ghat}
\end{equation}
Parametrizing the momentum space using the coordinates $p = p^\tau$, $\varphi_p$ and $v^z = \tanh(y - \eta)$, the Boltzmann Eq.~\eqref{eq:boltz_y} becomes
\begin{multline}
 \frac{\partial f}{\partial \tau} + \mathbf{v}_\perp \cdot \boldsymbol{\nabla} f - \frac{p^\tau v_z^2}{\tau} \frac{\partial f}{\partial p^\tau} - \frac{v^z(1 - v_z^2)}{\tau} \frac{\partial f}{\partial v^z} \\
 = -\frac{v^\mu u_\mu}{T_{\rm ref} L_{\rm ref}} \hat{\gamma} T (f - f_{\rm eq}),
 \label{eq:boltz_vz_aux}
\end{multline}
with $v^\mu = p^\mu / p^\tau = (1, \mathbf{v}_\perp, v^z/\tau)$, while $\mathbf{v}_\perp = \sqrt{1 - v_z^2}(\mathbf{i} \cos\varphi_p + \mathbf{j} \sin\varphi_p)$. Using the reduced distribution $\mathcal{F}$ from Eq.~\eqref{eq:F_red} and expressing everything in a dimensionless way, we arrive at
\begin{multline}
 \frac{\partial \overline{\mathcal{F}}}{\partial \bar{\tau}} + \mathbf{v}_\perp \cdot \overline{\boldsymbol{\nabla}} \overline{\mathcal{F}} + \frac{1 + v_z^2}{\bar{\tau}} \overline{\mathcal{F}} - \frac{\partial}{\partial v^z} \left[\frac{v^z(1 - v_z^2)}{\bar{\tau}} \overline{\mathcal{F}}\right] \\
 = -v^\mu u_\mu \hat{\gamma} \overline{T} (\overline{\mathcal{F}} - \overline{\mathcal{F}}_{\rm eq}).
 \label{eq:boltz_reduced}
\end{multline}
The bar quantities are nondimensionalized with respect to the reference quantities introduced in Eq.~\eqref{eq:ref}:
\begin{equation}
 \bar{\tau} = \frac{\tau}{L_{\rm ref}}, \quad 
 \overline{\boldsymbol{\nabla}} = L_{\rm ref} \boldsymbol{\nabla}, \quad 
 \overline{T} = \frac{T}{T_{\rm ref}}, \quad 
 \overline{\mathcal{F}} = \frac{\mathcal{F}}{\epsilon_{\rm ref}}.
\end{equation}
Equation~\eqref{eq:boltz_reduced} is solved using the Relativistic Lattice Boltzmann (RLB) solver described in Ref.~\cite[Sec. IV.B]{Ambrus:2021fej}, employing the Gauss-Legendre quadrature of order $Q_z = 80$ to discretize $v^z$ and $Q_{\varphi_p} = 40$ equidistant values of $\varphi_p$. The simulation domain in the transverse plane spans $[-L,L] \times [-L,L]$, with $L = 8R$ determined separately for each event, which is discretized using $100 \times 100$ equally-sized cells, which is sufficient since the solver employs a fifth-order advection scheme \cite{Ambrus:2021fej,Ambrus:2022koq,Ambrus:2024hks,Ambrus:2024eqa}. The initial time is set to $\tau_0 = 10^{-6} R$ and the distribution is initialized using the Romatschke-Strickland distribution~\cite{Romatschke:2003ms} with anisotropy parameter $\xi = 40$, corresponding to a ratio $P_L /P_T \simeq 0.04$.

\section{Momentum triangularity}
\label{sec:momentum_triangularity}

To obtain an estimate for the momentum triangularity coefficient $v_3$, we employ the first-order Chapman-Enskog approximation for the distribution function,
\begin{equation}
 f = f^{(0)} + f^{(1)}, \quad 
 f^{(0)} = \frac{1}{e^{p^\mu u_\mu/T} - 1},
\end{equation}
with $f^{(0)}$ corresponding to local thermal equilibrium. The correction $f^{(1)}$ can be obtained to first order in the relaxation time $\tau_R$ by replacing $f \to f^{(0)}$ on the left-hand side of the Boltzmann Eq.~\eqref{eq:boltz},
\begin{multline}
 f^{(1)} = \frac{\tau_R}{T (p \cdot u)} \left[ p^{\langle \mu} p^{\nu \rangle} \sigma_{\mu\nu} + (p\cdot u) p^{\langle \mu \rangle} \left(D u_\mu - \frac{\nabla_\mu T}{T}\right) \right.\\
 \left. - (p \cdot u)^2 \left(\frac{DT}{T} + \frac{\theta}{3}\right) \right] f^{(0)} \tilde{f}^{(0)},
 \label{eq:f1_aux}
\end{multline}
where we used $\partial_\mu f^{(0)} = -f^{(0)} \tilde{f}^{(0)} p^\nu \partial_\mu(u_\nu / T)$, with $\tilde{f}^{(0)} = 1 + f^{(0)}$, as well as the orthogonal decomposition
\begin{multline}
 p^\mu p^\nu = p^{\langle \mu} p^{\nu \rangle} + (p \cdot u) (p^{\langle \mu \rangle} u^\nu + p^{\langle \nu \rangle} u^\mu) \\
 + (p \cdot u)^2\left( u^\mu u^\nu - \frac{1}{3} \Delta^{\mu\nu}\right).
\end{multline}

Neglecting the dissipative terms in the conservation Eqs.~\eqref{eq:hydro_cons} and using the conformal equation of state $\epsilon = 3P = aT^4$, we can derive
\begin{equation}
 DT + \frac{T}{3} \theta = 0,\quad T Du^\mu = \nabla^\mu T,
\end{equation}
such that Eq.~\eqref{eq:f1_aux} reduces to
\begin{equation}
 f^{(1)} \simeq \frac{5\pi_{\mu\nu}}{8 P T} \frac{p^{\langle \mu} p^{\nu \rangle}}{p \cdot u} f^{(0)} \tilde{f}^{(0)},
 \label{eq:f1}
\end{equation}
where we replaced $\sigma_{\mu\nu} \simeq \pi^{\mu\nu} / 2\eta$ by the first-order Navier-Stokes relation, obtained by neglecting all second-order terms in Eq.~\eqref{eq:hydro_pi}, and we used Eq.~\eqref{eq:tcoeffs} to eliminate the shear viscosity coefficient $\eta$. This replacement has the advantage that Eq.~\eqref{eq:f1} does not involve gradients and thus will not introduce further errors in the evaluation of $v_3$ due to the numerical evaluation of the shear tensor.

Writing the complex anisotropic flow coefficients $\mathcal{V}_n = v_n e^{i n \Psi_{p,n}}$, introduced in Eq.~\eqref{eq:vn_def}, as
\begin{equation}
 \mathcal{V}_n = \frac{\int_{\mathbf{x}_\perp} V_n(\mathbf{x}_\perp)}{\int_{\mathbf{x}_\perp} V_0(\mathbf{x}_\perp)}, \quad
 V_n = \frac{\nu_{\rm eff}}{(2\pi)^3} \int d^3p\, p_\perp e^{in \varphi_p} f,
\end{equation}
the functions $V_n = V_n^{(0)} + V_n^{(1)}$ receive an equilibrium and a dissipative correction, given by
\begin{align}
 V_n^{(i)} &= \frac{\nu_{\rm eff}}{(2\pi)^3} \int dp\, p^3 \int_0^{2\pi} d\varphi_p \int_{-1}^1 dv^z\,\sqrt{1 - v_z^2} e^{i n \varphi_p} f^{(i)} \nonumber\\
 &= \int_0^{2\pi} d\varphi_p \int_{-1}^1 dv^z\,\sqrt{1 - v_z^2} e^{i n \varphi_p} \mathcal{F}^{(i)},
\end{align}
with $\mathcal{F}^{(0)}$ given in Eq.~\eqref{eq:Feq_red}, while 
\begin{align}
 \mathcal{F}^{(1)} &= \frac{\nu_{\rm eff}}{(2\pi)^3} \frac{5 \pi_{\mu\nu}}{8 P T} \frac{v^{\langle \mu} v^{\nu \rangle}}{v \cdot u} \int_0^\infty dp\, p^4 f^{(0)} \tilde{f}^{(0)} \nonumber\\
 &= \frac{5\pi_{\mu\nu}}{2P} \frac{v^{\langle \mu} v^{\nu \rangle}}{(v \cdot u)^2} \mathcal{F}^{(0)},
\end{align}
where we used the property $f^{(0)} \tilde{f}^{(0)} = -\frac{T}{v \cdot u} \partial_p f^{(0)}$, followed by an integration by parts. We obtain 
\begin{align}
 V_n^{(0)} &= \epsilon \int_0^{2\pi} \frac{d\varphi_p}{2\pi} e^{in \varphi_p} \int_{-1}^1 \frac{dv^z \sqrt{1 - v_z^2}}{2(v^\mu u_\mu)^4}, \nonumber\\
 V_n^{(1)} &= \frac{15}{2} \pi_{\mu\nu} \int_0^{2\pi} \frac{d\varphi_p}{2\pi} e^{in \varphi_p} \int_{-1}^1 \frac{dv^z \sqrt{1 - v_z^2}}{2(v^\mu u_\mu)^6} v^{\langle \mu} v^{\nu \rangle},
\end{align}
where $v^\mu u_\mu = \gamma (1 - \beta \sqrt{1 - v_z^2} \cos \Delta \varphi)$ for a boost-invariant flow configuration and $\Delta \varphi = \varphi_p - \varphi_u$. The case $n = 1$ can be evaluated exactly, due to the property $p_\perp e^{i \varphi_p} = p^x + i p^y$:
\begin{equation}
 V_1 = T^{0x} +i T^{0y}.
\end{equation}

To perform the azimuthal integral, we change the integration with respect to $\varphi_p$ to an integration with respect to $\Delta \varphi$ and write $e^{i n \varphi_p} = e^{i n \varphi_u} e^{i n \Delta \varphi}$. Introducing the notation
\begin{equation}
 I^{(n)}_k(a) = \int_0^{2\pi} \frac{d\varphi}{2\pi} \frac{e^{i n \varphi}}{(1 - a \cos\varphi)^k},
\end{equation}
we can express $V^{(i)}_n$ via
\begin{align}
 V^{(0)}_n &= \frac{\epsilon}{2\beta \gamma^4} e^{i n \varphi_u} \int_{-1}^1 dv^z a I^{(n)}_4, \nonumber\\
 V^{(1)}_n &= \frac{15}{4\beta\gamma^6} e^{in \varphi_u} \int_{-1}^1 dv^z a \nonumber\\
 &\times \Bigg\{\left[\left(2 - \frac{a^2}{\beta^2}\right) \pi_{00} - \left(1 - \frac{3a^2}{2\beta^2}\right) \pi_{(0)}\right] I^{(n)}_6 \nonumber\\
 & - \frac{a}{\beta} \sum_{m = \pm 1} e^{i m \varphi_u} \pi_{(-m)} I_6^{(n+m)} \nonumber\\
 &+ \frac{a^2}{4\beta^2} \sum_{m = \pm 2} e^{i m \varphi_u} \pi_{(-m)} I_6^{(n+m)}\Bigg\},
\end{align}
where the integrals $I^{(n)}_k$ take the argument $a = \beta \sqrt{1- v_z^2}$ and we introduced the irreducible combinations
\begin{gather}
 \pi_{(0)} = \pi^{xx} + \pi^{yy}, \quad 
 \pi_{(\pm 1)} = \pi^{0x} \pm i \pi^{0y}, \nonumber\\
 \pi_{(\pm 2)} = \pi^{xx} - \pi^{yy} \pm 2i \pi^{xy}.
 \label{eq:picomps}
\end{gather}
In deriving the expression for $V^{(1)}_n$, we took into account that the terms odd in $v^z$ vanish under the $v^z$ integration. We also used the tracelessness condition $\pi^\mu{}_\mu = 0$ to express $\tau^2 \pi^{\eta\eta} = \pi^{00} - \pi^{xx} - \pi^{yy}$. 

Using the following recurrence relations:
\begin{equation}
 a I^{(n)}_k = I^{(n-1)}_k - \frac{n + k - 2}{k - 1} I^{(n-1)}_{k-1}, \quad 
 I^{(0)}_{k+1} = I^{(0)}_k + \frac{a}{k} \frac{dI_k}{da},
\end{equation}
together with the initial values
\begin{equation}
 I^{(n)}_0 = \delta_{n0}, \quad 
 I^{(0)}_1 = \frac{1}{\sqrt{1 - a^2}} = \frac{\gamma}{\sqrt{1 + \beta^2 \gamma^2 v_z^2}},
\end{equation}
the required $I^{(n)}_k$ integrals can be obtained. In the case of $V^{(0)}_n$, we require
\begin{align}
 I^{(0)}_4 &= \frac{1 + \frac{3}{2}a^2}{(1 - a^2)^{7/2}}, &
 I^{(1)}_4 &= \frac{a(2 + \frac{1}{2} a^2)}{(1 - a^2)^{7/2}}, \nonumber\\
 I^{(2)}_4 &= \frac{5a^2}{2(1 - a^2)^{7/2}}, &
 I^{(3)}_4 &= \frac{5a^3}{2(1 - a^2)^{7/2}},
\end{align}
which leads to
\begin{subequations}
\begin{align}
 V^{(0)}_0 &= \frac{\epsilon \gamma^3}{\beta} \int_{0}^1 dv^z \frac{a + \frac{3}{2} a^3}{(1 + \beta^2 \gamma^2 v_z^2)^{7/2}}, \\
 V^{(0)}_1 &= \frac{\epsilon \gamma^3}{2\beta} e^{i \varphi_u} \int_0^1 dv^z \frac{a^2(4 + a^2)}{(1 + \beta^2 \gamma^2 v_z^2)^{7/2}}, \\
 V^{(0)}_2 &= \frac{5 \epsilon \gamma^3}{2\beta} e^{2i \varphi_u} \int_0^1 dv^z \frac{a^3}{(1 + \beta^2 \gamma^2 v_z^2)^{7/2}}, \\
 V^{(0)}_3 &= \frac{5 \epsilon \gamma^3}{2\beta} e^{3i \varphi_u} \int_0^1 dv^z \frac{a^4}{(1 + \beta^2 \gamma^2 v_z^2)^{7/2}},
\end{align}
\end{subequations}
where we took into account that $1/\sqrt{1 - a^2} = \gamma / \sqrt{1 + \beta^2 \gamma^2 v_z^2}$.
For $V^{(1)}_n$, the following integrals are required:
\begin{align}
 I^{(0)}_6 &= \frac{1 + 5a^2 + \frac{15}{8} a^4}{(1-a^2)^{11/2}}, & I^{(1)}_6 &= \frac{3a(1 + \frac{3}{2} a^2 + \frac{1}{8} a^4)}{(1 - a^2)^{11/2}}, \nonumber\\
 I^{(2)}_6 &= \frac{21a^2(1 + \frac{1}{2} a^2)}{4(1 - a^2)^{11/2}}, & 
 I^{(3)}_6 &= \frac{7a^3(1 + \frac{1}{8}a^2)}{(1 - a^2)^{11/2}}, \nonumber\\
 I^{(4)}_6 &= \frac{63a^4}{8(1 - a^2)^{11/2}}, & 
 I^{(5)}_6 &= \frac{63 a^5}{8(1 - a^2)^{11/2}},
\end{align}
together with the property $I^{(-n)}_k = I^{(n)}_k$ for the case when $n < 0$. Similar, but lengthier expressions can be obtained for $V^{(1)}_n$. 

To facilitate the $v^z$ integration, we introduce the following functions:
\begin{align}
 J_{n,m}(\beta) &= \int_0^1 dt (1 - t^2)^{n/2} (1 - \beta^2 t^2)^{m/2}  \nonumber\\
 &= \frac{\gamma}{\beta^n} \int_0^1 dv^z \frac{a^n}{(1 + \beta^2 \gamma^2 v_z^2)^{(m + n + 3)/2}},
\end{align}
where $t$ is related to $v^z$ via
\begin{equation}
 t = \frac{\gamma v^z}{\sqrt{1 + \beta^2 \gamma^2 v_z^2}}, \quad 
 v^z = \frac{t}{\gamma\sqrt{1 - \beta^2 t^2}}.
\end{equation}
In terms of the $J_{n,m}$ functions, $V^{(0)}_n$ becomes
\begin{align}
 V^{(0)}_0 &= \epsilon \gamma^2 \left(J_{1,3} + \frac{3}{2} \beta^2 J_{3,1}\right), \nonumber\\
 V^{(0)}_1 &= \frac{\epsilon \beta \gamma^2}{2} e^{i\varphi_u} (4 J_{2,2} + \beta^2 J_{4,0}), \nonumber\\
 V^{(0)}_2 &= \frac{5 \epsilon \beta^2 \gamma^2}{2} e^{2i\varphi_u} J_{3,1}, \nonumber\\
 V^{(0)}_3 &= \frac{5\epsilon \beta^3 \gamma^2}{2} e^{3i\varphi_u} J_{4,0}.
\end{align}

Using the recurrence relation
\begin{align}
 J_{n,m}(\beta) &= \frac{n J_{n-2,m}(\beta) + m  J_{n,m-2}(\beta)}{n+m+1}, 
 \label{eq:Jnm_rec_double}
\end{align}
valid for $n,m > 0$, as well as 
\begin{align}
 J_{n,m}(\beta) &= (1 - \beta^2) J_{n,m-2}(\beta) + \beta^2J_{n+2,m-2}(\beta),
\end{align}
we can derive the following single-index recurrences:
\begin{align}
 \beta^2 J_{n+2,m}(\beta) &= \frac{(2n+m+3) \beta^2 - n -1}{n+m+3} J_{n,m}(\beta) \nonumber\\
 &+ \frac{n(1 - \beta^2)}{n+m+3} J_{n-2,m}(\beta), \nonumber\\
 J_{n,m+2}(\beta) &= \frac{n + 2m + 3 - (m + 1) \beta^2}{n + m + 3} J_{n,m}(\beta) \nonumber\\
 & - \frac{m(1 - \beta^2)}{n + m + 3} J_{n,m-2}(\beta).
 \label{eq:Jnm_rec_single}
\end{align}
To find the functions $J_{n,m}$ required for the computation of $V^{(0)}_n$ and $V^{(1)}_n$, Eq.~\eqref{eq:Jnm_rec_double} is applied whenever both $n$ and $m$ are larger than $0$. Once either $n$ or $m$ has reached $0$ or $-1$, we apply the recurrence relations in Eq.~\eqref{eq:Jnm_rec_single} to decrease the value of the other index. When both $n$ and $m$ are even integers, the integrand in $J_{n,m}$ becomes a polynomial in $t$ and the recurrence is started with 
\begin{equation}
 J_{0,0}(\beta) = 1, \quad J_{0,2}(\beta) = 1 - \frac{1}{3} \beta^2, \quad 
 J_{2,0}(\beta) = \frac{2}{3}.
\end{equation}
When $n$ and $m$ are odd integers, the functions $J_{n,m}$ can be expressed with respect to the complete elliptic integrals of the first (${\rm K}$) and second (${\rm E}$) kind, defined as \cite{NIST:DLMF}:
\begin{align}
 {\rm K}(\beta) &= \int_0^1 \frac{dt}{\sqrt{(1-t^2)(1-\beta^2 t^2)}}, \nonumber\\
 {\rm E}(\beta) &= \int_0^1 dt \frac{\sqrt{1 -\beta^2 t^2}}{\sqrt{1 - t^2}}.
 \label{eq:elliptic_def}
\end{align}
In this case, the recurrence is started with
\begin{gather}
 J_{-1,-1}(\beta) = {\rm K}(\beta), \quad 
 J_{-1,1}(\beta) = {\rm E}(\beta), \nonumber\\
 J_{1,-1}(\beta) = \frac{1}{\beta^2} {\rm E}(\beta) - \frac{1-\beta^2}{\beta^2} {\rm K}(\beta).
\end{gather}

Applying the above recurrence relations, the functions $J_{n,m}$ required for the computation of $V^{(0)}_n$ can be readily derived:
\begin{align}
 J_{2,2}(\beta) &= \frac{2}{3} - \frac{2\beta^2}{15},\quad 
 J_{4,0}(\beta) = \frac{8}{15}, \nonumber\\
 J_{1,3}(\beta) &= \frac{3 + 7 \beta^2 -2\beta^4}{15\beta^2} {\rm E}(\beta) - \frac{3 + \beta^2}{15 \beta^2 \gamma^2} {\rm K}(\beta), \nonumber\\
 J_{3,1}(\beta) &= -\frac{2-7\beta^2 - 3\beta^4}{15\beta^4} {\rm E}(\beta) + \frac{2(1 - 3\beta^2)}{15\beta^4 \gamma^2} {\rm K}(\beta).
\end{align}
This leads to the following expressions for $V^{(0)}_n$:
\begin{subequations}
\begin{align}
 V^{(0)}_0 &= \frac{\epsilon}{6} [(7 + \beta^2) \gamma^2 {\rm E} - 4 {\rm K}], \\
 V^{(0)}_1 &= \frac{4 \gamma \epsilon}{3} e^{i \varphi_u} \beta \gamma, \\
 V^{(0)}_2 &= -\frac{\epsilon e^{2i \varphi_u}}{3 \beta^2} [\left(1 - \tfrac{7}{2}\beta^2 - \tfrac{3}{2}\beta^4\right) \gamma^2 {\rm E}
  - (1 - 3 \beta^2) {\rm K}], \\
 V^{(0)}_3 &= \frac{4 \gamma \epsilon}{3} e^{3i \varphi_u} \beta^3 \gamma,
\end{align}
\end{subequations}
where the elliptic functions take the argument $\beta$, i.e. ${\rm E} \equiv {\rm E}(\beta)$ and ${\rm K} \equiv {\rm K}(\beta)$.
It is easy to recognize that 
\begin{equation}
 V^{(0)}_1 = T^{0x}_{(0)} + i T^{0y}_{(0)}, \quad 
 V^{(0)}_3 = (T^{0x}_{(0)} + i T^{0y}_{(0)}) \beta^2 e^{2i \varphi_u}.
\end{equation}

Performing the same steps for $V^{(1)}_n$, we arrive at
\begin{subequations}
\begin{align}
 V^{(1)}_0 &= \frac{1}{16 \beta^2 \gamma^2} \Big\{\pi_{(0)} [(3 + 7 \beta^2 - 2 \beta^4) \gamma^2 {\rm E} - (3 + \beta^2) {\rm K}] \nonumber\\ 
 & - \frac{\pi^{00}}{\beta^2}  [(6 - 15 \beta^2 + 5 \beta^4 - 4 \beta^6) \gamma^2 {\rm E} \nonumber\\
 &- 2(3 - 6 \beta^2 + \beta^4) {\rm K}]\Big\}, \\
 V^{(1)}_1 &= \pi^{0x} + i \pi^{0y}, \\
 V^{(1)}_2 &= \frac{1}{32 \gamma^2} \left\{ \frac{3\pi_{(2)}}{2} \left[(29 - 9 \beta^2 + 4\beta^4) \gamma^2 {\rm E} - 2(7 - \beta^2) {\rm K}\right] \right.\nonumber\\
 & - \frac{e^{2i \varphi_u} \pi_{(0)}}{\beta^2} [(10 - 17 \beta^2 + 27 \beta^4 - 12 \beta^6) \gamma^2 {\rm E} \nonumber\\
 &- 2(5 - 6 \beta^2 + 3 \beta^4) {\rm K}] \nonumber\\
 & - \frac{e^{4i \varphi_u} \pi_{-2}}{2\beta^4} \left[(48 - 112 \beta^2 + 65 \beta^4 + 3\beta^6 - 12 \beta^8) \gamma^2 {\rm E} \right. \nonumber\\
 & \left.\left. - 2(24 -44 \beta^2 + 15 \beta^4 + 3\beta^6){\rm K}\right]\right\},\\
 V^{(1)}_3 &= \frac{1}{2} [\beta e^{i \varphi_u} \pi_{(2)} (6 - 3\beta^2 + \beta^4) -
 \beta^3 e^{3i \varphi_u} \pi_{(0)} (5 - 2 \beta^2) \nonumber\\
 &+ \beta^5 e^{5i \varphi_u}\pi_{(-2)} ],
\end{align}
\end{subequations}
where again the elliptic functions take the argument $\beta$, i.e. ${\rm E} \equiv {\rm E}(\beta)$ and ${\rm K} \equiv {\rm K}(\beta)$.

\section{Event-by-event spread}\label{app:spread}

\begin{figure*}
     \centering
    \includegraphics[width=.46\linewidth]{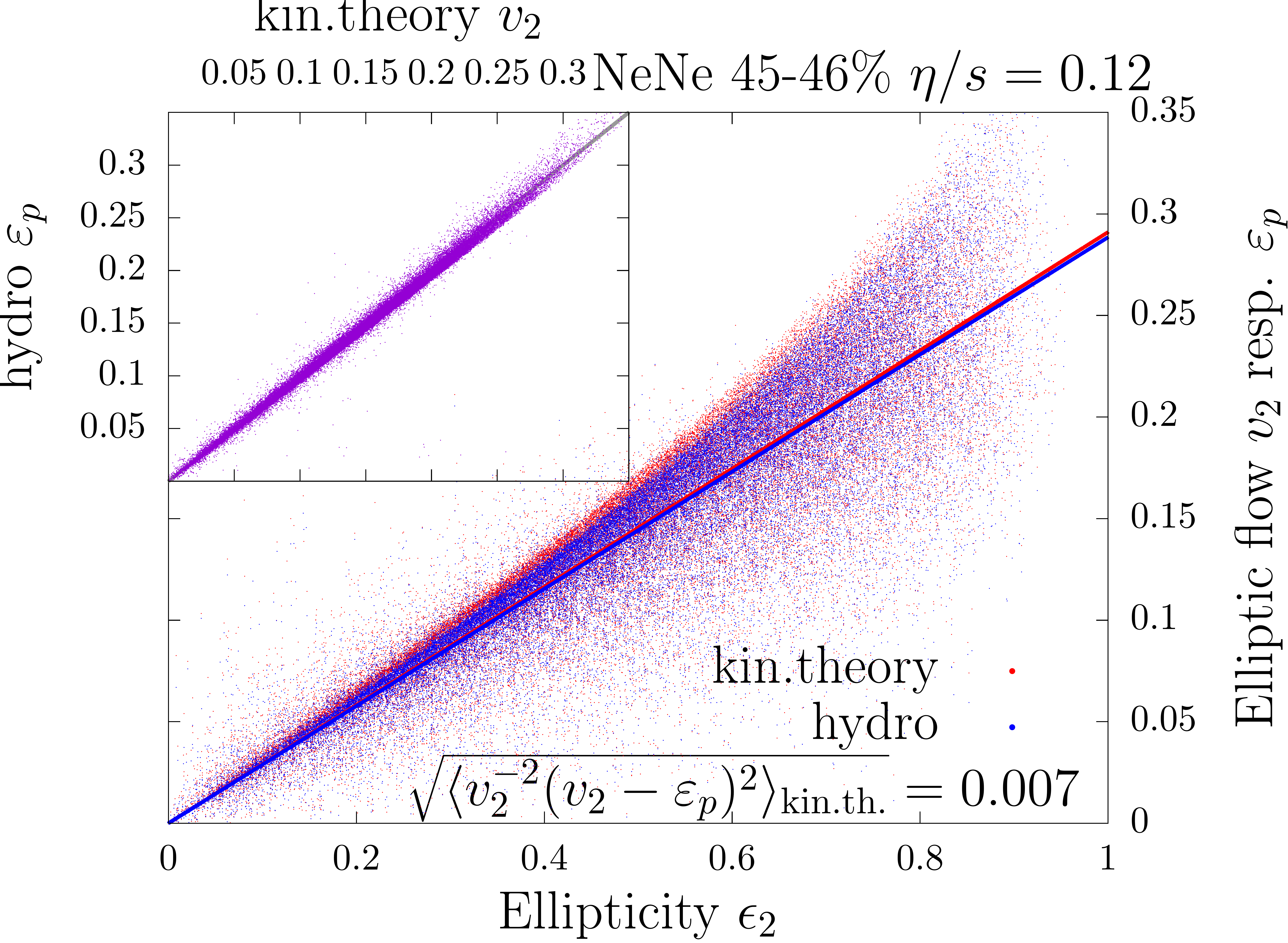}\hspace{20pt}
     \includegraphics[width=.46\linewidth]{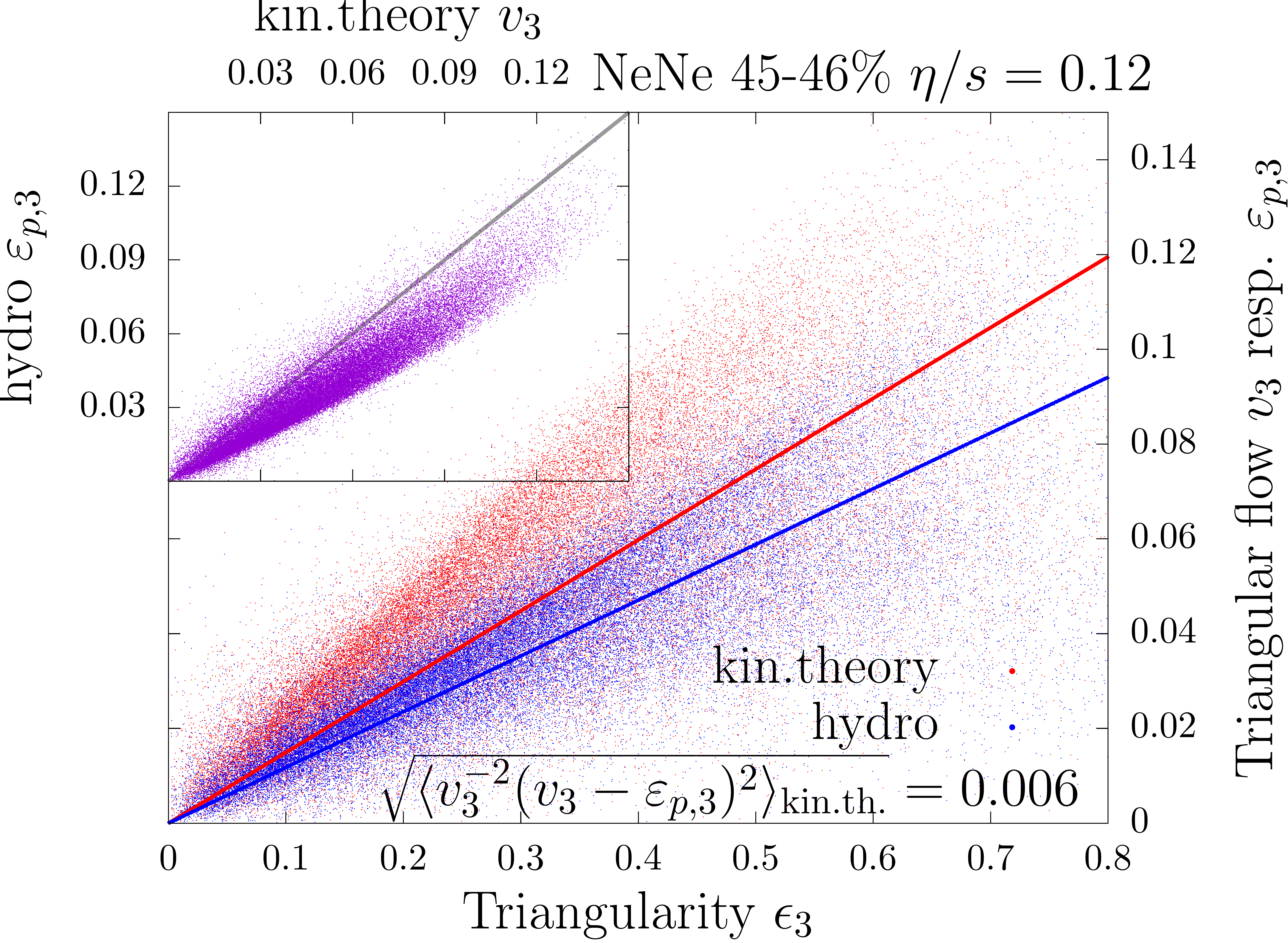}
     \caption{Event-by-event results for (left) elliptic flow versus ellipticity and (right) triangular flow versus triangularity from kinetic theory as $v_n$ (red) and hydrodynamics as $\varepsilon_{p(,3)}$ (blue) from the 45-46\% centrality class of Ne+Ne collisions at $\eta/s=0.12$. Lines in the main panels are linear fits to the scatter plots.}
     \label{fig:ebeepvsv45}
 \end{figure*}

\begin{figure*}
     \centering
    \includegraphics[width=.46\linewidth]{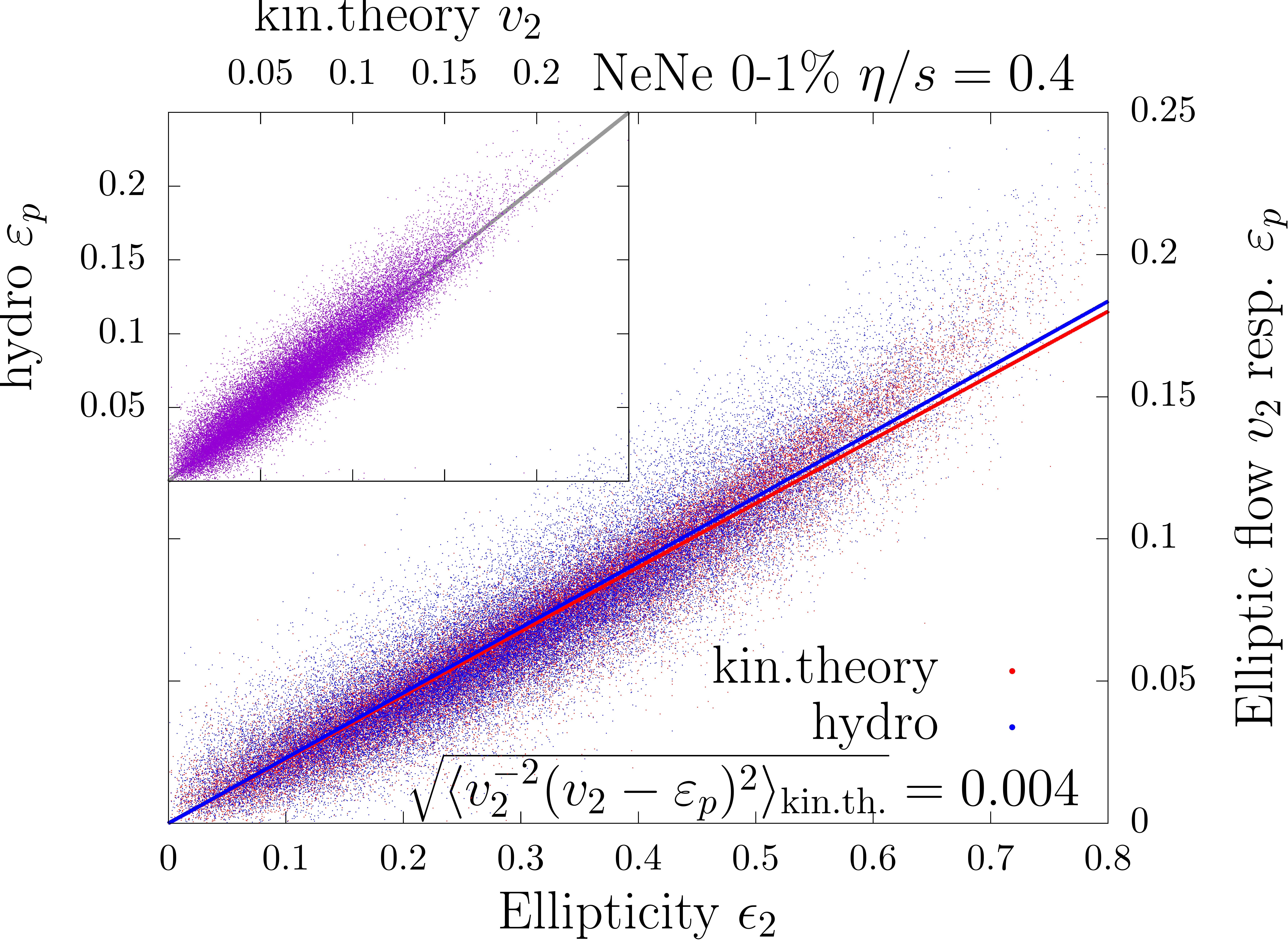}\hspace{20pt}
     \includegraphics[width=.46\linewidth]{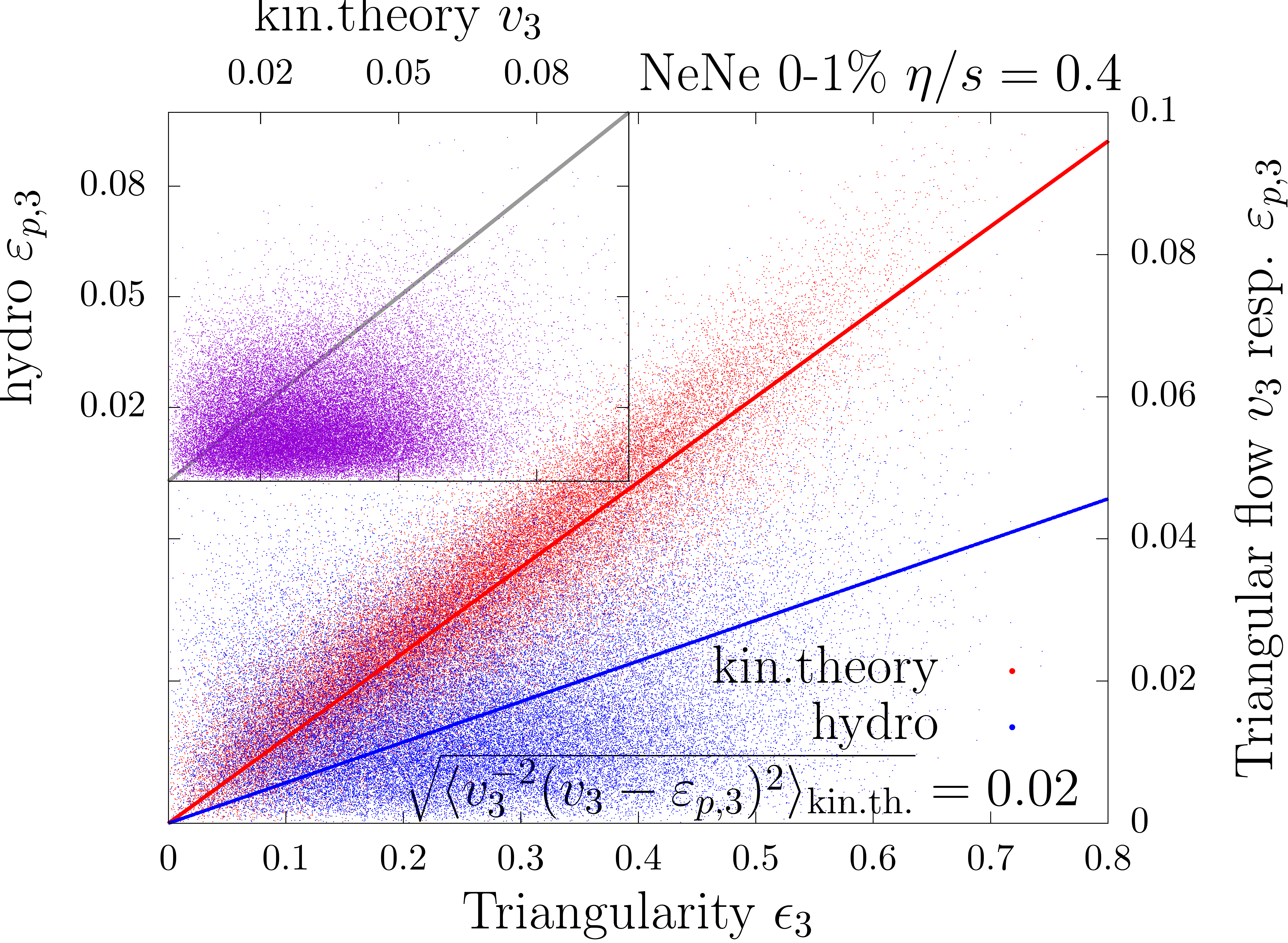}
     \caption{Same as Fig.~\ref{fig:ebeepvsv45}, but for the 0-1\% centrality class of Ne+Ne collisions at $\eta/s=0.4$.}
     \label{fig:ebeepvsvl}
 \end{figure*}

\begin{figure*}
     \centering
    \includegraphics[width=.46\linewidth]{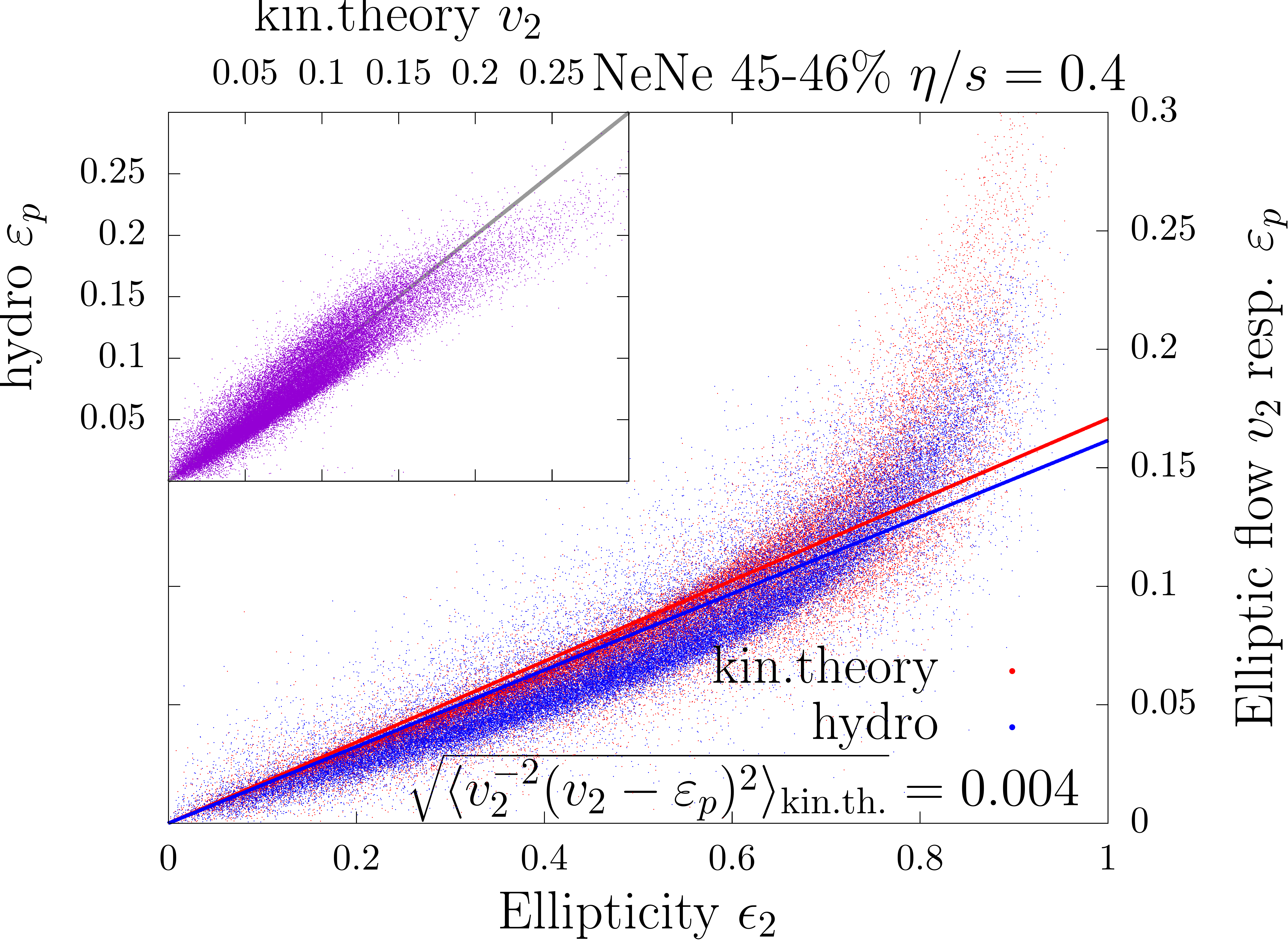}\hspace{20pt}
     \includegraphics[width=.46\linewidth]{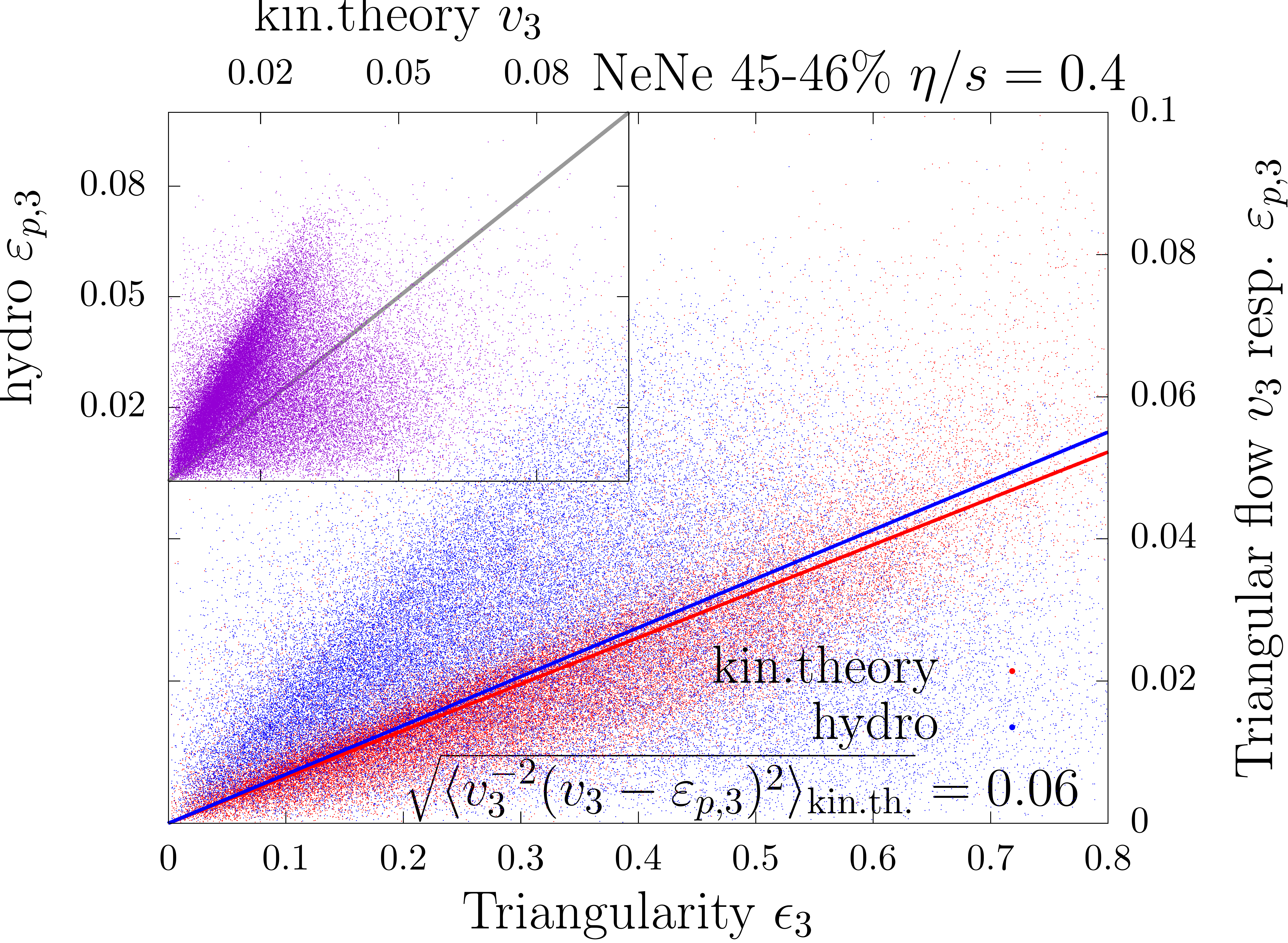}
     \caption{Same as Fig.~\ref{fig:ebeepvsv45}, but for the 45-46\% centrality class of Ne+Ne collisions at $\eta/s=0.4$.}
     \label{fig:ebeepvsv45l}
 \end{figure*}

In Figs.~\ref{fig:ebeepvsv45}--\ref{fig:ebeepvsv45l}, we show additional plots similar to the ones in Fig.~\ref{fig:ebeepvsv} but for the 45-46\% centrality class instead of 0-1\% or $\eta/s=0.4$ instead of $\eta/s=0.12$. The peripheral class shows a stronger spread towards larger eccentricities, indicating overall stronger eccentricity-nonlinear contributions to the flow response. At the same time, the spread of the comparison of hydro to kinetic theory becomes much larger. The latter effect also occurs at larger $\eta/s$. However, differences between the flow observables and their proxies remain small in kinetic theory, indicating that the extra spread comes from the breakdown of hydrodynamics.

\clearpage

\bibliography{main}

\end{document}